\documentclass[%
 twocolumn, 
 showkeys,
superscriptaddress,
 amsmath,
 amssymb,
 aps, 
prb,
]{revtex4-2}

\usepackage{amsmath,amssymb,graphicx,url,mathtools,hyperref,caption,physics,xcolor}
\hypersetup{hidelinks}
\usepackage{booktabs}
\usepackage{siunitx}
\usepackage{amsmath}
\usepackage{placeins}

\begin{document}


\title{The Frequency-Dependent Spin Contribution to the Magnetoelectric Tensor of Cr$_2$O$_3$: A First-Principles Study}



\author{Torsten Geirsson}
\affiliation{Institute of Materials Science (ICMUV), University of Valencia, Catedrático Beltrán 2, E-46980 Valencia, Spain}
\email[Contact author: ]{torsten.geirsson@outlook.com}
\author{Davide Sangalli}
\affiliation{Istituto di Struttura della Materia-CNR (ISM-CNR)  and European Theoretical Spectroscopy Facility (ETSF), Piazza Leonardo da Vinci 32, 20133 Milano, Italy}
\author{Alberto García-Cristóbal}
\affiliation{Institute of Materials Science (ICMUV), University of Valencia, Catedrático Beltrán 2, E-46980 Valencia, Spain}
\author{Alejandro Molina-Sánchez}
\affiliation{Institute of Materials Science (ICMUV), University of Valencia, Catedrático Beltrán 2, E-46980 Valencia, Spain}


\date{\today}

\begin{abstract}
The magnetoelectric (ME) effect provides a promising pathway for controlling magnetic functionalities using electric fields. While first-principles approaches to the static linear ME response have been rigorously developed and successfully benchmarked, comparable methods for the frequency-dependent linear ME effect remain far less established. This is despite numerous experimental studies demonstrating pronounced resonance effects associated with the ME response at finite frequencies. In this work, we investigate the dynamical spin-induced linear ME response from first-principles and provide a systematic comparison of different theoretical frameworks. We implement and assess the independent-particle approximation (IPA), random-phase approximation (RPA), time-dependent density functional theory (TDDFT), and the Bethe–Salpeter equation (BSE) to evaluate their performance in describing the frequency-dependent spin ME tensor. These methods are applied to the prototypical ME material Cr$_2$O$_3$, and the results are compared to available experimental and theoretical studies. We find that the IPA- and RPA-level descriptions fail to reproduce the previously reported finite static limit of the spin-induced ME response. Within the BSE framework, pronounced excitonic resonances emerge in the ME spectrum, in qualitative agreement with experiment. In addition, a magnon-like peak is identified, coinciding with a pole of the transverse spin susceptibility while remaining essentially dark in the optical absorption spectrum, highlighting the sensitivity of the ME response to spin excitations. TDDFT yields the magnon mode closer to the expected low-energy magnonic regime and produces a sizable static spin-induced ME response. Our results demonstrate that different theoretical frameworks capture complementary aspects of the dynamical ME response: an accurate description of low-energy collective spin excitations is required to recover the static limit, while electron–hole interactions are essential to reproduce the excitonic resonances of the ME spectrum.
\end{abstract}


\maketitle


\section{\label{section:introduction} Introduction}
The magnetoelectric (ME) effect describes a coupling between magnetic and electric degrees of freedom in solids. It refers to the induction of electric polarization in response to an external magnetic field, or conversely, an induced magnetization in the presence of an external electric field. Although seemingly distinct, these two phenomena are thermodynamically related in the static equilibrium limit, with the former often referred to as the direct effect and the latter as the converse effect. The strength of this coupling is quantified by the so-called ME tensor~\cite{newnham2004properties}. 

This coupling offers a route toward electrical control of magnetic properties, with prospective applications including low-power memory, spintronic logic, and ultrafast magneto-optical devices. In this context, ME coupling is often discussed within the broader class of multiferroic materials, in which magnetic and electric order coexist and interact~\cite{spaldin2005renaissance,spaldin2019advances}. The technological relevance of the ME effect depends critically on the magnitude of the coupling, which is typically too small in most single-phase crystals for direct device applications. Composite heterostructures, on the other hand, can exhibit significantly enhanced effective ME responses, leading to a range of proposed device concepts~\cite{liang2021roadmap}. Despite their more limited application potential, single-phase magnetoelectrics remain essential model systems for elucidating the intrinsic microscopic mechanisms that govern ME coupling. Among these, the prototypical magnetoelectric Cr$_2$O$_3$ (chromia) has played a central role due to its well-characterized crystal and magnetic structure, and its suitability for first-principles investigations. 

Early measurements of the ME effect focused primarily on responses driven by static or quasi-static external fields. More recent optical and terahertz pump-probe experiments, however, have revealed pronounced ME signatures at finite frequencies, including optical rotation and ellipticity associated with excitonic~\cite{krichevtsov1996magnetoelectric} and magnetic resonances~\cite{bilyk2025control}. These observations demonstrate that the ME response can be strongly enhanced at intrinsic excitation energies. Dynamical ME effects have also been investigated using ac measurement techniques, where the ME voltage coefficient can act as a sensitive probe of magnetic phase transitions~\cite{he2025dynamic}. Together, these studies highlight the importance of understanding the ME response beyond the static limit.

From a theoretical perspective, substantial progress has been made in the first-principles description of the static linear ME response. In particular, the modern theory of polarization has provided a rigorous framework for computing the ME tensor within periodic density-functional theory, enabling controlled calculations based on finite electric fields and allowing the response to be decomposed into electronic and lattice-mediated contributions~\cite{malashevich2012full,malashevich2010theory,ye2014dynamical}. 

By contrast, theoretical descriptions of the frequency-dependent ME response remain relatively limited, particularly at the first-principles level. While model and phenomenological approaches have provided valuable insight into magneto-optical and spin–light coupling mechanisms, a systematic understanding of how electronic excitations contribute to the dynamical ME tensor is still lacking. In this work, we present a first-principles study of the frequency-dependent electron spin-mediated ME tensor and analyze how different microscopic excitations contribute to the dynamical ME response. By comparing the independent-particle approximation (IPA), random phase approximation (RPA), time-dependent density functional theory (TDDFT), and the Bethe–Salpeter equation (BSE), we show how interband transitions, magnons, and excitons shape the ME spectrum. Using Cr$_2$O$_3$ as a benchmark system, we identify the origin of the low-energy magnonic contribution to the static ME response and the excitonic resonances observed in magnetoelectric spectroscopy experiments. In particular, we show that the dynamical ME response separates into distinct physical regimes: the static and low-frequency response is governed by collective spin excitations (magnons), while the optical response is dominated by electron-hole correlations in the excitonic regime. These contributions are captured at different levels of theory, highlighting the complementary roles of TDDFT and BSE in describing the electronic spin contribution to the frequency-dependent ME response.

\section{\label{section:theoretical_background} Theoretical background}
\subsection{Linear magnetoelectric effect in Cr$_2$O$_3$}
The linear ME effect is described by an axial second-rank tensor, the ME tensor, which we denote by $\alpha_{\mu\nu}^\mathcal{M}$ if the induced effect is magnetization $\mathbf{M}$ in the presence of an electric field $\boldsymbol{\mathcal{E}}$, and $\alpha_{\mu\nu}^\mathcal{P}$ if the induced effect is polarization $\mathbf{P}$ in the presence of a magnetic field $\boldsymbol{\mathcal{B}}$. In the static equilibrium, dissipationless limit, they are related by
\begin{equation}
    \alpha^\mathcal{M}_{\mu\nu} = \frac{\partial M_\nu}{\partial\mathcal{E}_\mu}\bigg|_{\mathcal{E}=\mathbf{0}} = \frac{\partial P_\mu}{\partial \mathcal{B}_\nu}\bigg|_{\boldsymbol{\mathcal{B}}=\mathbf{0}} = \alpha^\mathcal{P}_{\mu\nu}.
    \label{eq:static_me}
\end{equation}
This equality follows from the Maxwell relation $\partial^2 E/\partial \mathcal{B}_\nu\partial \mathcal{E}_\mu=\partial^2 E/\partial \mathcal{E}_\mu\partial \mathcal{B}_\nu$ where $E$ is the electromagnetic enthalpy density $E=E_0-\boldsymbol{\mathcal{E}}\cdot \mathbf{P} - \boldsymbol{\mathcal{B}}\cdot \mathbf{M}$~\cite{vanderbilt2018berry,essin2010orbital}. Here, we use the convention that the first index labels the electric-field/polarization direction, while the second index labels the magnetic-field/magnetization direction. The ME tensor is often decomposed into a frozen-ion contribution and a lattice-mediated contribution. For $\alpha_{\mu\nu}^{\mathcal{M}}$, both contributions can further be decomposed into orbital and spin components~\cite{malashevich2010theory}. The lattice-mediated contribution is expected to be largest in the static limit or when the frequency of the external field matches a phonon resonance. At optical frequencies, however, away from phonon resonances, the response is expected to be dominated by the purely electronic contribution, with the coupling mechanism between the spin degrees of freedom and an external electric field made possible through spin--orbit coupling (SOC). The ME response at these frequencies has been probed in spectroscopy experiments~\cite{krichevtsov1996magnetoelectric}, motivating a first-principles treatment beyond the static limit. 

While most computational studies have focused on the induced magnetization from a static external field, i.e. the static ME tensor defined in Equation~\eqref{eq:static_me}, the frequency-dependent ME tensor encodes rich information about how electric and magnetic excitations interact with optical fields~\cite{bilyk2025control}. Given that the ME response is typically weak~\cite{brown1968upper}, it can be enhanced if the frequency of an external field resonates with intrinsic excitations of the system, such as phonons, magnons, and excitons~\cite{fiebig2004revival,vopson2017measurement}. The dynamical ME response considered here arises from the linear response of the equilibrium electronic system to time-dependent fields, and is distinct from current-induced magnetization or spin–orbit torque phenomena~\cite{ying2022magnetoelectricity}.

The static definition of the ME tensor can be readily extended to finite frequencies:
\begin{equation}
    \begin{split}
    &\alpha^\mathcal{M}_{\mu\nu}(\omega) = \frac{\delta M_\nu(\omega)}{\delta\mathcal{E}_\mu(\omega)}, \hspace{1cm}
     \alpha^\mathcal{P}_{\mu\nu}(\omega) = \frac{\delta P_\mu(\omega)}{\delta \mathcal{B}_\nu(\omega)}.
    \end{split}
\end{equation}
These tensors are no longer guaranteed to be equal by the same thermodynamic argument used in the static case, since the thermodynamic potential $E$ exists only in the static equilibrium limit. In this work, we neglect any ionic or orbital contributions and focus on the spin magnetization change in response to an external electric field at finite frequencies. In other words, we focus on the part of $\alpha_{\mu\nu}^\mathcal{M}$ arising only from changes in spin magnetization under the clamped-ion approximation. We denote this purely electronic spin contribution to the ME response as $\alpha^\text{spin}_{\mu\nu}(\omega)$. Previous computational studies on Cr$_2$O$_3$~\cite{malashevich2012full, bousquet2011unexpectedly} suggest that the electronic spin contribution accounts for nearly the entire static electronic ME response, including both spin and orbital contributions, and for approximately 20--30\% of the total static ME response, including both electronic and lattice contributions. First-principles studies on the frequency dependence of the ME tensor remain scarce, even though substantial ME effects have been observed at excitonic resonance energies of Cr$_2$O$_3$~\cite{krichevtsov1996magnetoelectric,hayashida2022observation}. A microscopic theory based on localized electrons on each Cr$^{3+}$ has been developed to explain this phenomenon~\cite{muto1998magnetoelectric}, in which the ME response is reproduced through a perturbative treatment of crystal-field splittings and spin--orbit coupling, but it does not constitute a first-principles framework.

\subsection{Ab initio independent-particle ME tensor}
At the IP level, the magnetic response $\delta \mathbf{M}$ of an insulating material to an external electric field $\boldsymbol{\mathcal{E}}(\omega)$ is obtained within first-order perturbation theory by considering the coupling to the external field in the length gauge \mbox{$\mathcal{H}^\text{int}(\omega) = \boldsymbol{\mathcal{E}}(\omega)\cdot (-e\mathbf{r})$}~\cite{kaxiras2019quantum,rzkazewski2004equivalence}:
\begin{equation}
\begin{split}
    \alpha_{\mu\nu}(\omega) &= \frac{\mu_Bea_0}{VN_k\hbar}\sum_{vc\mathbf{k}}\left( \frac{R_{\mu\nu}^{vc\mathbf{k}}}{\omega-(\omega_{cv\mathbf{k}}-i\eta/\hbar)}\right. \\
    &\hspace{3cm}\left.-\frac{\left(R_{\mu\nu}^{vc\mathbf{k}}\right)^*}{\omega+(\omega_{cv\mathbf{k}}+i\eta/\hbar)}\right),
\label{eq:MEspin_IP_resp}
\end{split}
\end{equation}
where $\omega_{cv\mathbf{k}}=(\varepsilon_{c\mathbf{k}}-\varepsilon_{v\mathbf{k}})/\hbar$ are the frequency differences between conduction and valence states, $V$ is the unit cell volume, $N_k$ is the number of sampled k-points in the first Brillouin zone, and $\eta$ is a broadening factor. The IP residuals are given by
\begin{equation}
    R_{\mu\nu}^{vc\mathbf{k}} = \bra{v\mathbf{k}}m_\nu\ket{c\mathbf{k}}\bra{v\mathbf{k}}(-er_\mu)\ket{c\mathbf{k}}^*/(\mu_Bea_0),
    \label{eq:MEspin_IP_res}
\end{equation}
where $m_\nu$ is the $\nu$-component of the magnetic dipole operator, and $r_\mu$ is the $\mu$-component of the position operator. We use the magnitude of the electric charge \mbox{$e>0$}, the Bohr magneton $\mu_B$ and the Bohr radius $a_0$ to write the residuals in dimensionless form. For the electronic spin contribution $\alpha_{\mu\nu}^\text{spin}(\omega)$, the magnetic dipole operator is given by \mbox{$\mathbf{m}=-\mu_B\boldsymbol{\sigma}$}, where $\boldsymbol{\sigma}$ is the vector of Pauli matrices~\cite{pozo2023multipole}. If inversion symmetry is present, residuals are odd under $\mathbf{k}\to-\mathbf{k}$ and the Brillouin-zone integral vanishes at all frequencies. On the other hand, if time-reversal symmetry is present, $\mathcal{T}[R_{\mu\nu}^{vc\mathbf{k}}]=-(R_{\mu\nu}^{vc(-\mathbf{k})})^*$, which does not in general force the finite-frequency mixed response to vanish in non-centrosymmetric systems. In the limit \mbox{$\omega,\eta \to 0$}, however, the contributions from time-reversed partners cancel, consistent with the absence of a static linear magnetoelectric effect in time-reversal-symmetric materials. Furthermore, for a collinear magnetic configuration without spin--orbit coupling, the Kohn--Sham Hamiltonian is block diagonal in spin and the Bloch states can be chosen with a well-defined spin character. Since the electric dipole operator is spin independent, spin-flip transitions have vanishing electric-dipole matrix elements, whereas spin-conserving transitions have vanishing interband matrix elements of the spin magnetic dipole by orthogonality. Consequently, the independent-particle spin contribution vanishes unless SOC mixes the spin sectors.

\subsection{BSE, TDDFT and RPA approaches}
\label{sec:BSE_TDDFT_RPA}
We account for electron-hole interactions by solving the BSE in the form of an eigenvalue problem~\cite{marsili2021spinorial,wu2022optical}:
\begin{equation}
\begin{split}
    &\left(\varepsilon_{c\mathbf{k}}-\varepsilon_{v\mathbf{k}}\right)A^\lambda_{vc\mathbf{k}} + \sum_{v'c'\mathbf{k}'}\bra{vc\mathbf{k}}K\ket{v'c'\mathbf{k}'}A^\lambda_{v'c'\mathbf{k}'} \\
    &\hspace{6cm}= E^\lambda A^\lambda_{vc\mathbf{k}}
\end{split}
\label{eq:BSE_eigenvalue}
\end{equation}
where $\varepsilon_{n\mathbf{k}}$ are now quasiparticle energies, $K$ is the BSE kernel, and $A^\lambda_{vc\mathbf{k}}$ are the exciton coordinates in the basis of free electron-hole pairs of the exciton state \mbox{$\ket{\lambda} = \sum_{vc\mathbf{k}}A^\lambda_{vc\mathbf{k}}\ket{vc\mathbf{k}}$} with energy $E_\lambda$. We employ the Tamm--Dancoff approximation, which neglects the coupling between resonant and anti-resonant transitions and renders the BSE kernel Hermitian, thereby facilitating the solution of the eigenvalue problem. This approximation has been shown to be reliable in the optical regime for a wide range of materials~\cite{martin2016interacting}. For the ME response, the Bethe--Salpeter eigenvectors mix the formerly independent transitions in \eqref{eq:MEspin_IP_resp} to form the BSE ME residuals $R_{\mu\nu}^\lambda$:
\begin{equation}
\begin{split}
    R_{\mu\nu}^{\lambda} &= \frac{1}{N_k\mu_Bea_0}\left(\sum_{vc\mathbf{k}}\bra{v\mathbf{k}}m_\nu\ket{c\mathbf{k}}A^\lambda_{vc\mathbf{k}}\right) \\ 
    &\hspace{2.0cm}\times \left(\sum_{vc\mathbf{k}}\bra{v\mathbf{k}}(-er_\mu)\ket{c\mathbf{k}}A^\lambda_{vc\mathbf{k}}\right)^*,
    \label{eq:MEspin_BSE_res}
\end{split}
\end{equation}
where the first term in parentheses is the magnetic dipole matrix element $\bra{0}m_\nu\ket{\lambda}$ between the ground-state and a given exciton state, and the second term in parentheses is the electric dipole matrix element $\bra{0}(-er_\mu)\ket{\lambda}$, with $\ket{0}$ being the many-body ground-state. The ME tensor then becomes
\begin{equation}
\begin{split}
    \alpha_{\mu\nu}(\omega) &= \frac{\mu_Bea_0}{V\hbar}\sum_{\lambda}\left(\frac{R_{\mu\nu}^\lambda}{\omega-(E_\lambda-i\eta)/\hbar}\right. \\
    &\hspace{3cm}\left.-\frac{\left(R_{\mu\nu}^\lambda\right)^*}{\omega+(E_\lambda+i\eta)/\hbar}\right). \\
\end{split}
\label{MEspin_BSE_resp}
\end{equation}
As they stand, Eqs. \eqref{eq:MEspin_IP_resp} and \eqref{MEspin_BSE_resp} are in the units of A/V. They can either be converted to a dimensionless susceptibility by multiplying the expression by the impedance of free space $Z_0 = 1/\varepsilon_0c$~\cite{muto1998magnetoelectric} or to s/m by multiplying the tensor with the magnetic permeability of the material. In this work we follow the latter convention unless stated otherwise. We approximate the magnetic permeability for Cr$_2$O$_3$ by that of vacuum, which is an adequate approximation for our purposes~\cite{malashevich2012full}.

The BSE kernel $K$ in Equation~\eqref{eq:BSE_eigenvalue} is defined as \mbox{$K=v+F^\text{BSE}$}, where the Hartree term $v$ and screened interaction $F^\text{BSE}$ are four-point functions in space, spin and time $1\equiv(\mathbf{r}_1,\sigma_1,t_1)$:
\begin{subequations}
\begin{equation}
    v(1,2,3,4) = \delta(1,3)\delta(2,4)v(1,2)
\end{equation}
\begin{equation}
    F^\text{BSE}(1,2,3,4) = -\delta(2,4)\delta(1,3)W(1,2)
\end{equation}
\end{subequations}
and $W(1,2)=\int\varepsilon^{-1}(1,3)v(3,2)\text{d}[3]$ is the screened interaction obtained from the dielectric function $\varepsilon(1,2)$ and bare Coulomb interaction $v$. While the BSE provides a description of the neutral excitations through the Hartree and screened interaction terms, it can be written in an analogous linear-response Dyson-like form as in the time-dependent density functional theory (TDDFT) formalism, differing only in the structure of their interaction kernel~\cite{reining2002excitonic}. In TDDFT, the interaction is captured by the Hartree term and the two-point exchange-correlation kernel, \mbox{$K = v+f_{xc}$}. An essential difference between the BSE and TDDFT approaches is that TDDFT is formulated in terms of Kohn--Sham (KS) eigenvalues and orbitals, whereas the BSE is built on quasiparticle (QP) eigenvalues and orbitals. In practical implementations of the BSE, however, the KS orbitals are often used as approximations for the QP orbitals, while the eigenvalues are replaced by QP-corrected energies. Furthermore, the calculation of the screened interaction $W$ makes the BSE approach computationally demanding. This is particularly true for two-dimensional systems, where the screened interaction converges slowly~\cite{guandalini2023efficient}. 

In this work, we use adiabatic time-dependent density-functional theory (A-TDDFT), for which the exchange-correlation kernel is frequency independent. In practice, we employ the adiabatic local-density approximation (ALDA) for the kernel~\footnote{For both the LDA and the GGA ground state, in the \textsc{Lumen} code, only the term containing the functional derivative with respect to the density is constructed in the definition of the adiabatic kernel, i.e. $f^A_{xc}=\partial v^{DFT}[\rho_{\alpha\beta}(t)](\mathbf{x},t)/\partial \rho_{\alpha\beta}(\mathbf{x'},t)$.}, which we refer to as TD-ALDA in the following. Within TD-ALDA, the $f_{\rm xc}$ is local in space and therefore lacks the long-range, non-local, and frequency-dependent structure required to reproduce bound excitonic features in extended systems. Nevertheless, TD-ALDA enables us to assess the role of the long-range electron-hole attraction, which is included in the BSE through the screened interaction.
Finally, we also consider the RPA, which amounts to retaining the Hartree term $v$ in the kernel. Since SOC is essential for the spin-mediated ME response, we use a spinorial formulation of BSE and A-TDDFT as described in Ref.~\cite{marsili2021spinorial}. Together, these formulations provide a unified framework for analyzing the spin ME response from independent-particle transitions to correlated excitonic excitations.

\section{Computational details}

\subsection{Cr$_2$O$_3$ as a prototypical magnetoelectric}

\begin{figure}[t]
\begin{center}
\includegraphics[scale=0.45]{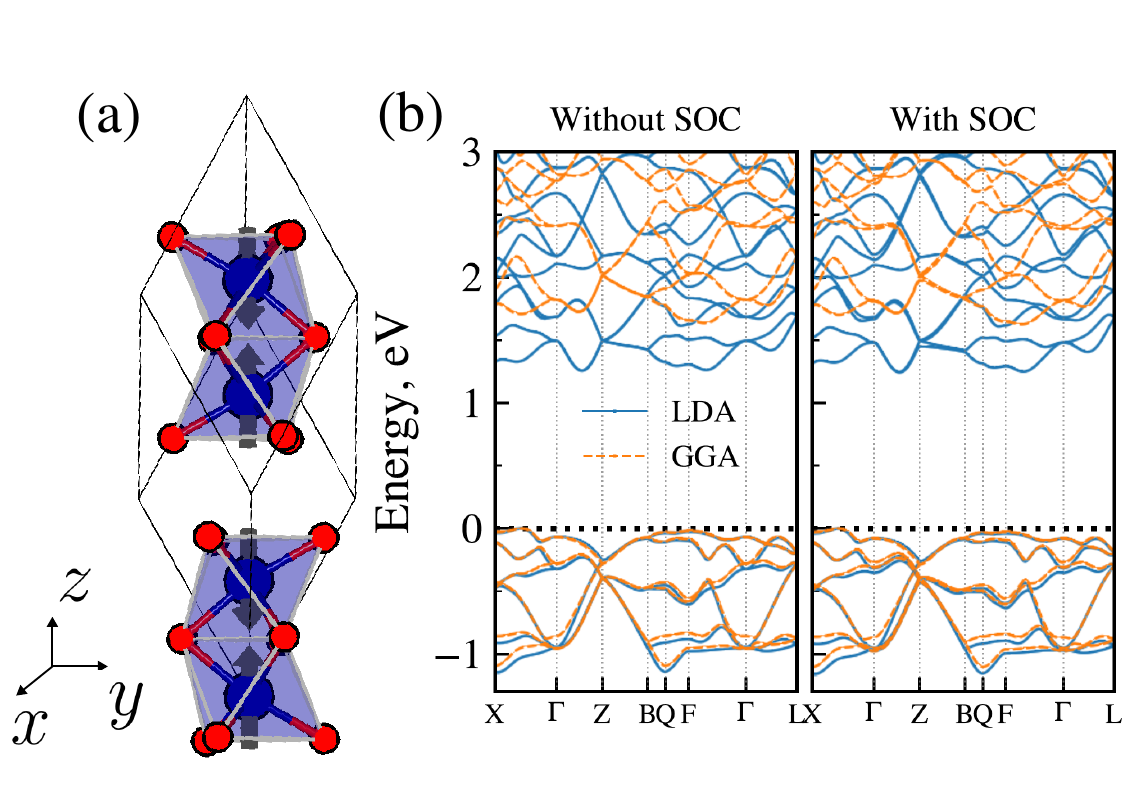}
    \caption{(a) The unit cell geometry and electronic band structure of Cr$_2$O$_3$. The grey arrows indicate the direction of the magnetic moments of the Cr$^{3+}$ ions. (b) The bands are obtained from DFT ground-state electronic calculations using LDA and GGA exchange-correlation functionals, with and without spin--orbit coupling (SOC). The energy zero is set at the energy of the highest occupied state for each set of bands. Note that they are doubly degenerate due to the combined time-reversal and inversion symmetry of Cr$_2$O$_3$.}
    \label{fig:bands+geometry}
\end{center}
\end{figure}
In this work, we develop and assess first-principles computational frameworks for computing the frequency-dependent electron spin-mediated ME tensor $\alpha_{\mu\nu}^\text{spin}(\omega)$. We apply these methods to the insulating antiferromagnet Cr$_2$O$_3$, whose magnetic point group is $\bar{3}'m'$.
 
The structural optimization and ground state electronic calculations were carried out with the plane-wave pseudopotential DFT code \textsc{Quantum ESPRESSO}~\cite{QE-2017}.
The geometry was optimized using a spin-polarized collinear configuration without SOC. The GGA (Perdew-Burke-Ernzerhof~\cite{PBE1996}) exchange-correlation functional was used in combination with optimized norm-conserving Vanderbilt pseudopotentials~\cite{Hamann2013_ONCVPSP}. The k-mesh used for the optimization was an $8\times 8\times 8$ $\Gamma$-centered Monkhorst-Pack grid. The wavefunction and charge density cutoffs were 90 Ry and 360 Ry, respectively. The variable-cell geometry optimization was iterated until the force components on each atom were less than $10^{-3}$ Ry/$a_0$ and the cell pressure was less than 0.5 kbar. The final rhombohedral lattice parameter was $a=5.420$ Å and rhombohedral angle $\alpha=54.52^\circ$. In comparison, the experimental values are $(a,\alpha) = (5.358\text{ Å}, 55.0^\circ)$~\cite{hill2010crystallographic}, in good agreement with our calculations. The atomic crystal coordinates are given in the Supplemental Material~\cite{si}, and the geometry is displayed in Figure~\ref{fig:bands+geometry}(a). 

In order to investigate the sensitivity of the ME tensor to the exchange-correlation functional, both LDA (Perdew-Zunger~\cite{PerdewZunger1981}) and GGA were tested, with the self-consistent charge density recalculated in each case but with the geometry obtained with GGA. In Figure~\ref{fig:bands+geometry}(b), we display the DFT band structure obtained using the LDA and GGA exchange-correlation functionals, with and without SOC. The inclusion of SOC has little effect on the band energies, but the conduction band energies and band gap differ significantly between the two functionals. With SOC, LDA gives a band gap of 1.3~eV, which is 0.4~eV lower than the GGA band gap of  1.7~eV. Consequently, we calculate $\alpha_{\mu\nu}^\text{spin}(\omega)$ using both LDA and GGA exchange-correlation functionals in order to assess the sensitivity of the results to the underlying exchange-correlation functional.

\subsection{Calculation of the ME tensor}
The optimized geometry was used as input for non-collinear DFT+SOC calculations, which served as the ground state for all linear-response calculations. 
Due to the $\bar{3}'m'$ point group,~\cite{newnham2004properties} Cr$_2$O$_3$ exhibits a linear ME tensor composed of two independent components $\alpha = \text{diag}(\alpha_\perp,\alpha_\perp,\alpha_\parallel)$, where $\alpha_\perp=\alpha_{xx}=\alpha_{yy}$ is the transverse component and $\alpha_\parallel=\alpha_{zz}$ is the longitudinal component, in the coordinate system displayed in Figure~\ref{fig:bands+geometry}(a).

The calculations of the ME tensor and the dielectric tensor were carried out with the \textsc{Lumen} fork of \textsc{Yambo}~\cite{sangalli2019many,lumen2025}.  The main features of the IP/RPA/TD-ALDA dielectric and ME tensors were well converged with a $4\times 4\times 4$ k-mesh, which was shifted half a grid step in each direction. At IP/RPA level, 23 valence and 28 conduction bands were included. For TD-ALDA, the low-frequency structure and the static limit converged more slowly with the number of bands, and 40 valence and 40 conduction bands were required to obtain a reliable static limit, whereas the spectra between 1 and 5~eV were already well converged with only 20 valence and 20 conduction bands. The plane-wave cutoff in the expansion of the wavefunctions was 70 Ry for all IPA, RPA and TD-ALDA calculations.

The BSE calculations were performed using a coarse $6\times 6\times 6$ k-mesh and a finer k-mesh of $16\times 16\times 16$, according to the double grid method described in Ref.~\cite{alliati2022double}. The dielectric function and screened Coulomb interaction were calculated with a plane-wave cutoff of 12~Ry, and the exchange interaction with a 90 Ry cutoff. The dielectric function was computed using 500 bands. To converge the low-frequency structure of the ME tensor, the kernel was constructed from 12 valence bands and 28 conduction bands. Scissor shifts were applied to both the LDA and GGA band structures to place the direct QP gap at $E^{QP}_\text{gap}=3.4$~eV, consistent with the experimental gap inferred for Cr$_2$O$_3$ from combined XPS/BIS spectra~\cite{zimmermann1996electron} and commonly used as a reference in first-principles studies~\cite{lebreau2014structural,shi2009magnetism}. We used a broadening of $\eta=0.1$~eV in the ME response function unless otherwise specified. For the evaluation of the static limit $\alpha^\text{spin}(0)$, however, a smaller broadening of $\eta = 0.01$~eV was used to obtain a well-converged estimate.

\section{\label{section:results_and_discussion} Results and Discussion}

In this section we compare the IPA, RPA, TD-ALDA and BSE approaches, clarifying how each level of theory captures microscopic excitations that shape the finite-frequency ME response, such as interband transitions, magnons, and excitons. By applying this methodology to Cr$_2$O$_3$, we (i) analyze excitonic and magnonic contributions to $\alpha^\text{spin}(\omega)$, (ii) identify the origin of the static ME response, and (iii) benchmark our results against computational and experimental studies of the static and frequency-dependent ME tensor available in the literature. In doing so, we provide a unified first-principles framework linking the dynamical ME tensor to specific elementary excitations.

Throughout this work we focus on the transverse component $\alpha_\perp^\text{spin}(\omega)$. We find that the longitudinal component $\alpha_\parallel^\text{spin}(\omega)$ is approximately an order of magnitude smaller than the transverse component $\alpha_\perp^\text{spin}(\omega)$ across all frequencies and levels of approximation considered. This pronounced anisotropy is consistent with previous computational studies of the static limit and has been attributed to the large stiffness of the magnitude of the spin magnetic moment in collinear antiferromagnets~\cite{malashevich2012full} at zero temperature. At higher temperatures, however, the parallel component exhibits large temperature variations~\cite{astrov1961magnetoelectric}, but since our calculations correspond to the zero-temperature limit, we focus entirely on $\alpha_\perp^\text{spin}(\omega)$. The parallel components $\alpha^\text{spin}_\parallel(\omega)$ obtained with TD-ALDA and BSE are plotted in the Supplemental Material~\cite{si} for completeness.

\begin{figure}[t]
\begin{center}
    \hspace{-0.2cm}\includegraphics[scale=1.0]{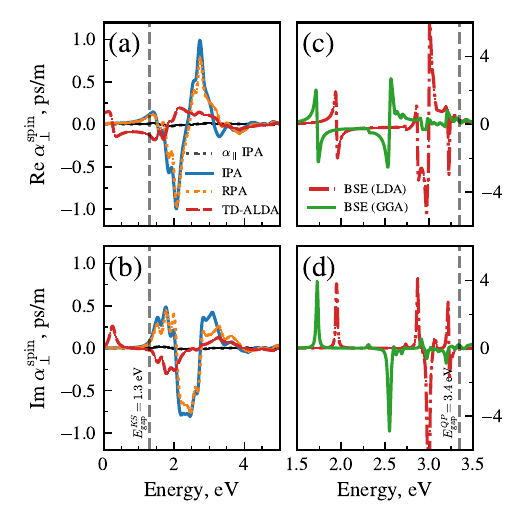}
\end{center}
\caption{ Panels (a) and (b): Real and imaginary parts, respectively, of the transverse component of the spin magnetoelectric tensor, $\alpha^{\rm spin}_\perp(\omega)$, computed within the independent-particle approximation (IPA), the random-phase approximation (RPA), and TD-ALDA, all from the LDA ground-state. The RPA kernel includes only the Hartree term which has a minor effect on the magnetoelectric tensor of Cr$_2$O$_3$ compared with the IPA result. The parallel component, $\alpha^{\rm spin}_\parallel(\omega)$, is also shown at the IPA level. Panels (c) and (d) show the corresponding spectra obtained from separate BSE calculations using LDA and GGA ground states. The vertical dashed lines indicate the Kohn--Sham and quasiparticle band gaps. The broadening in panels (a) and (b) is $\eta=0.1$~eV whereas we employ a smaller broadening of $\eta=0.01$~eV in panels (c) and (d) to focus on the excitonic region.}
\label{fig:ME_all}
\end{figure}

An overview of the results is provided in Figure~\ref{fig:ME_all}. The results reveal three distinct microscopic excitations that shape the frequency-dependent ME tensor: independent interband transitions, magnonic spin-flip excitations, and bound electron-hole excitations. As we show below, these contributions are captured differently by the different approximations. We discuss the IPA and RPA results in Sec.~\ref{sec:IP_RPA}, the TD-ALDA and BSE results in Sec.~\ref{sec:TD-ALDA_BSE}, the low-energy BSE states in
~\ref{sec:BSE-poles}, and the comparison with experiment in Sec.~\ref{sec:experimental}.

\subsection{IPA and RPA}
\label{sec:IP_RPA}
We first analyze the ME tensor in the simplest approximations, namely the IPA and RPA. In Figure~\ref{fig:ME_all}(a,b), the real and imaginary parts of $\alpha_\perp(\omega)$ are shown for these approximations. The qualitative shape is similar for the two approximations, demonstrating that inclusion of the Hartree term $v$ in the kernel has a small impact on the ME tensor in this system. The major features are in the range 1--4~eV where $\text{Im}\,\alpha_\perp(\omega)$ displays two clear changes in sign. Experimentally, sign reversals of the frequency-dependent ME tensor have been inferred from changes in rotation and ellipticity of reflected light across optical frequencies~\cite{krichevtsov1996magnetoelectric}. In those measurements, however, the observed features are associated with excitonic resonances. Since excitonic effects are not captured at the IPA and RPA levels, a quantitative comparison is only meaningful at the BSE level, which we discuss in the next section. In addition, a non-zero static limit of the ME tensor is neither reproduced in the IPA nor in the RPA, even though Cr$_2$O$_3$ is known to have a non-zero static spin ME response of approximately 0.3 ps/m~\cite{malashevich2012full,bousquet2011unexpectedly}. This demonstrates that independent transitions alone are insufficient to generate a static or low-frequency ME response in Cr$_2$O$_3$. The Kramers--Kronig relations imply that the static limit is governed by a frequency-weighted integral of the absorptive part, $\alpha^{\rm spin}_{\perp}(0)\propto \int d\omega\, {\rm Im}\,\alpha^{\rm spin}_{\perp}(\omega)/\omega$. A finite static response therefore requires low-energy spectral weight and/or an incomplete cancellation of positive and negative contributions, suggesting that collective spin excitations missing at the IPA/RPA level are essential. These effects are addressed in the following section using BSE and TD-ALDA approaches.

\subsection{TD-ALDA and BSE}
\label{sec:TD-ALDA_BSE}
\begin{figure}[t]
\begin{center}
    \includegraphics[scale=1.0]{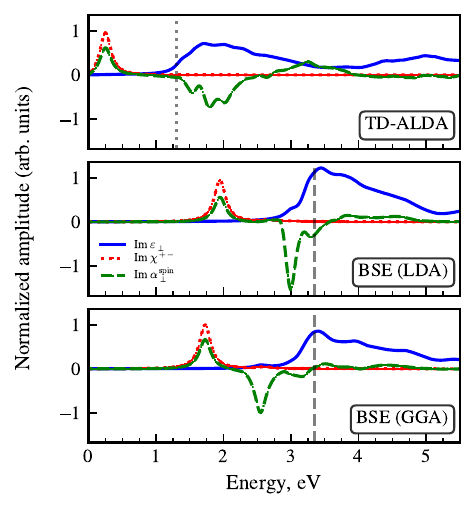}
\end{center}
    \caption{The imaginary parts of the transverse dielectric tensor component $\varepsilon_\perp$, spin susceptibility $\chi^{\pm}$, and transverse spin ME tensor component $\alpha_\perp^\text{spin}$ (all normalized with respect to their respective maximum value), obtained with TD-ALDA, and with the BSE using LDA and GGA ground-state electronic structures. The vertical dotted line indicates the Kohn--Sham band gap and the vertical dashed line indicates the quasiparticle band gap used in the BSE calculations. The lowest peak of the ME tensor coincides with the lowest peak of the spin susceptibility at vanishing optical absorption, whereas the second ME peak occurs at non-zero absorption but vanishing spin-susceptibility.}
\label{fig:BSE-TD-ALDA-overview}
\end{figure}
We next turn to the TD-ALDA and BSE results shown in Figure~\ref{fig:ME_all}. The shapes of $\alpha_\perp^\text{spin}(\omega)$ differ in significant ways compared to the IPA/RPA results, showing a rich structure at lower energies and clear sign changes. The results obtained with TD-ALDA show a low-frequency peak at 250 meV, which exhibits a slow convergence towards lower energies with respect to the number of bands included in the kernel. As we increased the number of bands and decreased the broadening value $\eta$ in Equation~\eqref{MEspin_BSE_resp}, the static limit of the TD-ALDA spectrum converged to a finite limit of approximately 0.2 ps/m, in reasonable agreement with the previously reported value of about 0.3 ps/m for the spin contribution to the static ME response~\cite{malashevich2012full,bousquet2011unexpectedly}. The sizable discrepancy between TD-ALDA and IPA/RPA highlights the importance of using an exchange–correlation kernel capable of capturing low–frequency magnonic excitations in order to recover a finite and physically meaningful static ME response. Furthermore, TD-ALDA is unable to capture any bound excitons, as expected due to the short-range nature of the kernel, $f_{xc}$~\cite{botti2005energy}.
\begin{figure}[t]
\begin{center}
\includegraphics[scale=1.0]{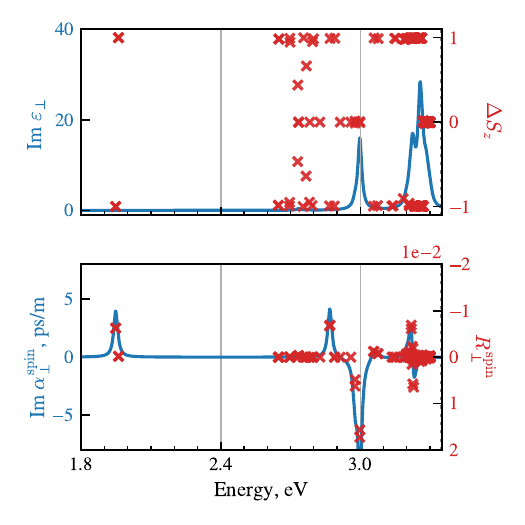}
\caption{The imaginary parts of the transverse dielectric (top) and ME (bottom) tensor components below the quasiparticle band gap at 3.4~eV, obtained with BSE (using the LDA ground-state). The spin characters $\Delta S_z$ are obtained by diagonalizing the excitonic spin operator within each degenerate exciton multiplet, and the BSE ME residuals $R^\text{spin}$ are obtained with Equation~\eqref{eq:MEspin_BSE_res}. Both $\Delta S_z$ and $R^\text{spin}$ are shown as red crosses. The four lowest states are doubly degenerate, with the two lowest being ME active and two highest being ME silent. The broadening is $\eta = 0.01$~eV. The residual axis is inverted to facilitate comparison with $\Im \alpha_\perp^\text{spin}.$}
\label{fig:BSE_oscillator-strengths}
\end{center}
\end{figure}
\begin{figure}[t]
\begin{center}
\hspace{-0.2cm}\includegraphics[scale=1.0]{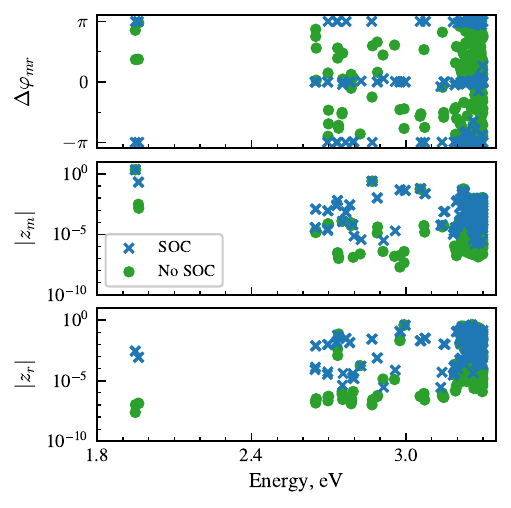}
\caption{The decomposition of the BSE (with LDA ground-state) residuals $R_\perp^\text{spin}=|z_m||z_r|e^{i\Delta\varphi_{mr}}$, with and without SOC. In both cases we use a non-collinear formalism. The results show that SOC locks the relative phase between the electric and magnetic transition matrix elements, leading to real-valued BSE ME residuals. In the absence of SOC, the electric and spin-magnetic transitions decouple and the ME response is strongly suppressed.}
\label{fig:BSE_oscillator-strengths_decomp}
\end{center}
\end{figure}

In order to characterize the nature of the lowest--energy TD-ALDA peak, we examine the transverse spin susceptibility $\text{Im}\,\chi^{\pm}(\omega)$ which provides direct information about spin--flip excitations~\cite{skovhus2022magnons,szilva2023quantitative}. As shown in Figure~\ref{fig:BSE-TD-ALDA-overview}, the low-energy feature in $\alpha_\perp^\text{spin}(\omega)$ lies in a region of vanishing optical absorption, but closely follows the peak of $\text{Im}\,\chi^{\pm}(\omega)$, indicating that the TD-ALDA peak originates from a spin-dominated excitation which is optically dark but has a strong ME character. This connection between the ME response and the spin susceptibility supports the interpretation that low–energy spin fluctuations play a central role in establishing the static limit of the ME tensor.

In contrast, the BSE spectrum does not display significant ME spectral weight at low energies, but instead at 1.9~eV with an LDA ground-state, and at 1.7~eV using a GGA ground-state. Consequently, the static limit is negligible within our numerical accuracy. However, by similarly computing $\text{Im}\,\chi^{\pm}(\omega)$, we see that the first peak shares the pole of the spin susceptibility, again indicating that the lowest ME peak originates from a spin-dominated excitation which is optically dark but ME active. Moreover, the excitonic region below the band-gap onset is characterized by comparatively weak optical absorption and small spin-susceptibility weight, yet it carries a comparatively strong ME response. This is consistent with both theoretical~\cite{muto1998magnetoelectric} and experimental~\cite{krichevtsov1996magnetoelectric} studies attributing the excitonic ME resonance to localized $d-d$ transitions of the Cr$^{3+}$ ions. We thus conclude that the ME response can reflect both excitonic and magnonic resonances in magnetic materials.

In order to analyze the BSE eigenstate contributions to the ME response, we plot the BSE ME residuals according to Equation~\eqref{eq:MEspin_BSE_res} along with the transverse ME function in Figure~\ref{fig:BSE_oscillator-strengths}. The lowest ME-active BSE pole corresponds to a cluster of two pairs of states, each pair being nearly degenerate and the pairs being separated by roughly $10$~meV. Given that Cr$_2$O$_3$ is antiferromagnetic and hosts four magnetic ions per unit cell, this quartet is naturally associated with the separately degenerate acoustic and optical modes at zero magnon momentum. These two observations strongly suggest that the lowest ME peak is a magnonic contribution to the dynamical ME tensor, whose absolute energy is misplaced due to the well-known Goldstone rule violation in $GW$--BSE approaches to magnons~\cite{esquembre2025magnons, olsen2021unified, muller2016acoustic}. Because the magnonic peak appears in the~eV range instead of in the meV range, its contribution to the static limit is suppressed by the $1/\omega$ factor in $\alpha_\perp^\text{spin}(0) \propto \int\text{Im}\,\alpha^\text{spin}_\perp(\omega)/\omega\,\text{d}\omega$. Furthermore, the mode-resolved residues show that only the two lowest magnonic branches carry a significant ME oscillator strength, while the two higher-energy states are relatively ME weak. A recent pump–probe experiment on Cr$_2$O$_3$ has directly demonstrated that a spin resonance can be excited by both electric and magnetic fields at 0.165~THz (0.682~meV), and that both driving mechanisms result in comparable spin dynamics. The oscillation amplitude scales linearly with the field strength and exhibits the expected sign reversal upon inversion of the magnetic Néel vector. Moreover, the authors conclude that the ME coefficient varies only weakly with frequency across the interval 0–0.2~THz, with the lowest electric-dipole-active phonon lying at a higher energy of 9.14~THz. The study therefore provides direct experimental evidence for a magnetoelectrically active magnon mode, consistent with the low-energy pole observed in our calculated dynamical ME tensor, which is intrinsically tied to the electronic static limit~\cite{bilyk2025control}.

To understand the origin of the sign changes in the ME response, we analyze the complex phase relation between electric and magnetic transition matrix elements. We decompose the BSE ME residuals in Equation~\eqref{eq:MEspin_BSE_res} according to
\begin{equation}
R^\text{spin}_\perp = z_mz_r^* = |z_m|\,|z_r|\,e^{i\Delta\varphi_{mr}}
\end{equation}
where \mbox{$z_m = \bra{0}m_x\ket{\lambda}/(\mu_B\sqrt{N_k})$} and \mbox{$z_r = \bra{0}-er_x\ket{\lambda}/(ea_0\sqrt{N_k})$}. The magnitude of the ME response of a given state $\lambda$ is then determined by the product $|z_m|\,|z_r|$, whereas the sign is determined by the phase difference $\Delta\varphi_{mr}$ between $z_m$ and $z_r$. These quantities are plotted in Figure~\ref{fig:BSE_oscillator-strengths_decomp}. In the presence of SOC, the phase differences are fixed by spin--orbit induced coupling between spin and orbital degrees of freedom. This results in real values of $R^\text{spin}_\perp$, with $\Delta\varphi_{mr}=\pm\pi$ yielding a positive value of $\text{Im}\,\alpha^\text{spin}_\perp(\omega)$ and $\Delta\varphi_{mr}=0$ yielding a negative value. This phase fixing disappears when SOC is switched off (while retaining non-collinearity), and electric and magnetic transitions decouple, resulting in unconstrained phase differences. In this case, the state-resolved residuals largely cancel when summed, strongly suppressing the ME response. This suppression is further enhanced by the reduced magnitudes of $|z_r|$ in the
absence of SOC.

\subsection{Microscopic character of the ME-active excitons}
\label{sec:BSE-poles}
To further clarify the microscopic nature of the BSE poles that dominate the magnetoelectric response, we analyze the spin character and orbital composition of the main ME-active excitons. The spin character $\Delta S_z$ is obtained by constructing the excitonic spin matrix and diagonalizing it within each nearly degenerate multiplet, following the approach of Ref.~\cite{kshirsagar2025flipping} as implemented in \textsc{Yambopy}~\cite{paleari_2025_15012963}, and further explained in the Supplemental Material. The values of $\Delta S_z$ are displayed in Figure~\ref{fig:BSE_oscillator-strengths}. Finally, the orbital character is estimated by combining the BSE amplitudes with atomic projections of the underlying Kohn--Sham states:
\begin{equation}
W_{\alpha\to\beta} = \frac{\sum_{vc\mathbf{k}} |A^{\lambda}_{vc\mathbf{k}}|^2 P^{\alpha}_{v\mathbf{k}} P^{\beta}_{c\mathbf{k}}}{\sum_{vc\mathbf{k}} |A^{\lambda}_{vc\mathbf{k}}|^2}
\end{equation}
where $P^\alpha_{n\mathbf{k}}$ is the projected weight of the KS state $\ket{n\mathbf{k}}$ onto the subspace of atomic orbitals $\alpha$. In the following, $W_{d\rightarrow d}$ refers specifically to the Cr-$(d)\rightarrow$Cr-$(d)$ transition weight, obtained by summing over all Cr-$(d)$-like projections.

The results obtained from the LDA ground-state are summarized in Table~\ref{table:me-microscopic}. All main ME-active poles have a dominant  Cr-$(d)\rightarrow$Cr-$(d)$ character, with \mbox{$W_{d\rightarrow d}\approx0.7$}. This supports the interpretation that the finite-frequency ME response in the excitonic region is mainly associated with Cr-$(d)$-dominated crystal-field-like excitations. The LDA analysis gives spin-flip character for the lowest-energy ME-active poles, with \(\Delta S_z=\pm1\), while the pole at 2.998~eV is predominantly spin conserving. This is broadly consistent with the experimental separation between spin-forbidden and spin-allowed $d-d$ features. However, as shown in the Supplemental Material, the precise spin character and ordering of individual poles depend on the electronic-structure starting point. We therefore avoid assigning individual BSE eigenstates uniquely to atomic multiplets. The robust conclusion is that both spin-flip and spin-conserving Cr-$(d)\rightarrow$ Cr-$(d)$ excitations can carry sizable ME oscillator strength.

\begin{table}
\caption{Character of the main ME-active BSE poles obtained from the LDA ground state. The energy and orbital weight are averaged over each nearly degenerate multiplet, while $R^{\mathrm{spin}}_{\perp}$ is summed over the states in the multiplet. The spin character \(\Delta S_z\) is obtained by diagonalizing the excitonic spin operator within each multiplet. Here \(W_{d\to d}\) denotes the Cr-$(d)\rightarrow$Cr-$(d)$ transition weight.}
\label{table:me-microscopic}

\begin{tabular}{c c c c c}
\hline
Energy (eV) & Deg. & $\Delta S_z$ &
$W_{d\to d}$ & $R_\perp^\text{spin}$ \\
\hline
1.947 & 2 & $\approx -1$ & $0.71$ & $-1.25\times 10^{-2}$ \\
1.960 & 2 & $\approx +1$ & $0.71$ & $-3.65\times 10^{-4}$ \\
2.869 & 2 & $\approx \pm1$ & $0.71$ & $-1.36\times 10^{-2}$ \\
2.998 & 2 & $\approx 0$ & $0.69$ & $+3.29\times 10^{-2}$ \\
\hline
\end{tabular}

\end{table}

\subsection{Comparison with optical experiments}
\label{sec:experimental}
\begin{figure}[b!]
\begin{center}
    \includegraphics[scale=1.00]{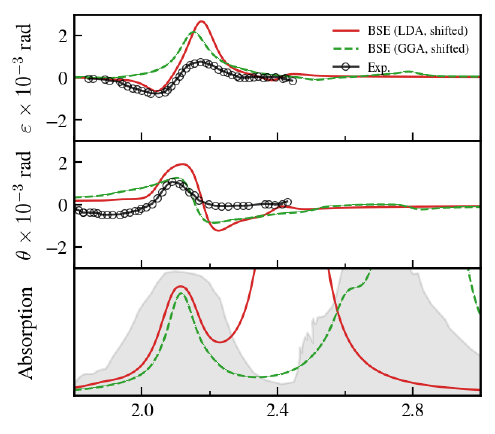}
\end{center}
\caption{Comparison between experimental rotation and ellipticity data from Ref.~\cite{krichevtsov1996magnetoelectric} and BSE results. The experimental rotation $\theta$ and ellipticity $\epsilon$ are multiplied by a constant factor of four to facilitate comparison of spectral shape and sign with theory. The BSE spectra are shifted downward in energy to align the first absorption feature with experiment, as explained in the text. In the lower panel, the red and green curves show the shifted BSE absorption spectra, while the gray shaded region shows the experimental absorption.}
\label{fig:experimental}
\end{figure}
In this section, we assess the predictive capability of BSE for the finite-frequency spin ME tensor. The finite-frequency ME response can be accessed experimentally via reflectance spectroscopy, where it gives rise to a non-reciprocal rotation and ellipticity of reflected light. Unlike the magneto-optical Kerr effect, this optical activity does not arise from a net magnetization of the material, but instead originates from the coupling between electric and magnetic degrees of freedom encoded in the ME tensor~\cite{hayashida2022observation}. The rotation $\theta$ and ellipticity $\epsilon$ of the polarization of light after reflection at the crystal surface are given by~\cite{hornreich1968theory,krichevtsov1996magnetoelectric}:
\begin{equation}
\theta + i\epsilon = \frac{2\alpha_\perp}{\varepsilon_0c}\,\frac{1+n_\perp}{1-n_\perp}
\label{eq:reflection}
\end{equation} 
where $n^2_\perp=\varepsilon_\perp$ is the square of the refractive index and $\alpha_\perp$ is in units of A$/$V. At low frequencies where the dispersion of $n_\perp$ is small, the rotation is mainly determined by Re $\alpha_\perp(\omega)$ and the ellipticity is mainly determined by Im $\alpha_\perp(\omega)$, and the experimental observables depend both on the amplitude as well as the sign of the ME tensor component. All quantities entering Equation~\eqref{eq:reflection} are available from the results obtained within any of the approximations considered, but since Ref.~\cite{krichevtsov1996magnetoelectric} measured the rotation and ellipticity at optical frequencies of 1.6--2.4~eV, which is in the excitonic regime, we compare their data with our BSE results. 

Figure~\ref{fig:experimental} shows the rotation and ellipticity obtained from Equation~\eqref{eq:reflection} using the BSE spectra. To align the absorption onset with the experimental data, the energy axis of the BSE spectra is shifted downward by 0.82~eV for the LDA ground-state and by 0.40~eV for the GGA ground-state. The ME-active excitonic peaks are known to be associated with both spin-forbidden and spin-allowed $d-d$ transitions. The discrepancy between theory and experiment may therefore partly reflect the tendency of LDA and GGA functionals to delocalize $d$-electrons, together with the approximation of QP energies by a scissor shift applied to the DFT band structure.

Nevertheless, the BSE results reproduce qualitative aspects of the experimental data, such as relative signs and the position and ordering of peaks in ellipticity and rotation with respect to the absorption peaks. The strongest ME activity is associated with the first absorption onset, whereas the second peak, with higher optical oscillator strength, is associated with lower ME activity. This is consistent with theoretical results obtained in Ref.~\cite{muto1998magnetoelectric} using a perturbative approach based on crystal field splitting and SOC. 

The experimental amplitudes are smaller by roughly a factor of four. This difference may arise from limitations of the electronic-structure description discussed above, contributions beyond the spin-electronic part, surface effects in the experimental samples, or other experimental uncertainties. To facilitate a direct comparison of spectral shape and sign, the experimental data in Figure~\ref{fig:experimental} are therefore multiplied by a constant factor of four. 

At the same time, clear differences are observed between the LDA and GGA results, most notably in the energy separation between the two absorption peaks, highlighting the sensitivity of the excitonic structure and the resulting ME response to the underlying electronic structure. A more accurate treatment of QP energies, for example through $GW$ corrections rather than a scissor shift, would likely improve the agreement between the two approximations. Overall, however, the comparison indicates that the BSE-based approach captures the essential excitonic contributions governing the optical ME response, while quantitative details remain dependent on the underlying electronic-structure description.

\section{Conclusions}
Our work presents a first-principles approach to computing the electronic spin contribution to the dynamical ME tensor in magnetic insulators and demonstrates how different levels of theory capture complementary aspects of the response. We find that the dynamical ME response is controlled by different physical mechanisms at low and high energies: collective spin excitations determine the static limit, while excitonic effects dominate the optical ME response. Overall, our results indicate that BSE, formulated in a spinorial framework and including SOC, provides a practical \textit{ab initio} route to the finite-frequency electronic ME response in the excitonic regime. A comparison with reflectance spectroscopy data shows qualitative agreement in the sign and spectral shape of the polarization rotation and ellipticity upon reflection. At lower frequencies, the predictive power of the present BSE-based approach is limited by its known violation of the Goldstone rule in the description of magnons. This leads to an unphysical placement of the low-energy magnonic pole and consequently prevents a reliable estimate of the static ME response. In this low-energy regime, a TDDFT framework appears more appropriate, as it provides a better placement of magnon energies and therefore yields a more reliable static limit, as confirmed by our TD-ALDA results. At energies in the meV range, additional contributions beyond the purely electronic spin term are expected to become important, most notably lattice-mediated mechanisms associated with infrared-active phonons. First-principles frameworks for the lattice-mediated static ME effect have been developed based on a decomposition of the ME response in terms of zone-center infrared phonon modes and provide a natural starting point for extending the present analysis toward the finite-frequency regime~\cite{iniguez2008first}. Furthermore, the orbital contribution to the ME tensor can be non-negligible in certain materials. The present work focuses on the spin-mediated response, which captures a sizable part of the purely electronic contribution in Cr$_2$O$_3$. An extension of the formalism would allow the inclusion of orbital contributions, which we leave for future work. Finally, the quantitative accuracy of the excitonic ME response depends sensitively on the underlying electronic structure, in particular on the description of localized $d$-electrons. While local and semi-local exchange--correlation functionals capture the qualitative features discussed here, more advanced treatments, such as DFT$+U$, hybrid functionals, or $GW$-based approaches, may be required for a fully predictive description of excitation energies and oscillator strengths in correlated magnetic materials. These aspects represent important directions for future work and will be essential for extending predictive ME modeling to a broader class of correlated magnetic materials.

\begin{acknowledgments}
This work is supported by the Horizon Europe research and innovation program of the European Union under the Marie Sklodowska-Curie grant agreement 101118915 (TIMES). This work is part of the project I+D+i PID2023-146181OB-I00 UTOPIA, funded by MCIN/AEI/10.13039/501100011033, the project PROMETEO/2024/4 (EXODOS). The authors gratefully acknowledge the computer resources at Agustina and technical support provided by BIFI and Barcelona Supercomputing Center (FI-2025-2-0001), and the computer resources at Tirant-UV (project lv48 - FI-2025-2-0001).
DS acknowledges funding from the ``MAterials design at the eXascale'' (MaX) centre of excellence co-funded by the European High Performance Computing joint Undertaking (JU) and participating countries (Grant Agreement No. 101093374), and from PRIN project ``Exploring extreme ultraviolet excitons with attosecond time resolution'' (EXATTO), Grant No. 2022PX279E from MIUR (Italy).
\end{acknowledgments}

\bibliography{main}

\providecommand{\noopsort}[1]{}\providecommand{\singleletter}[1]{#1}%
\begin{thebibliography}{53}%
\makeatletter
\providecommand \@ifxundefined [1]{%
 \@ifx{#1\undefined}
}%
\providecommand \@ifnum [1]{%
 \ifnum #1\expandafter \@firstoftwo
 \else \expandafter \@secondoftwo
 \fi
}%
\providecommand \@ifx [1]{%
 \ifx #1\expandafter \@firstoftwo
 \else \expandafter \@secondoftwo
 \fi
}%
\providecommand \natexlab [1]{#1}%
\providecommand \enquote  [1]{``#1''}%
\providecommand \bibnamefont  [1]{#1}%
\providecommand \bibfnamefont [1]{#1}%
\providecommand \citenamefont [1]{#1}%
\providecommand \href@noop [0]{\@secondoftwo}%
\providecommand \href [0]{\begingroup \@sanitize@url \@href}%
\providecommand \@href[1]{\@@startlink{#1}\@@href}%
\providecommand \@@href[1]{\endgroup#1\@@endlink}%
\providecommand \@sanitize@url [0]{\catcode `\\12\catcode `\$12\catcode
  `\&12\catcode `\#12\catcode `\^12\catcode `\_12\catcode `\%12\relax}%
\providecommand \@@startlink[1]{}%
\providecommand \@@endlink[0]{}%
\providecommand \url  [0]{\begingroup\@sanitize@url \@url }%
\providecommand \@url [1]{\endgroup\@href {#1}{\urlprefix }}%
\providecommand \urlprefix  [0]{URL }%
\providecommand \Eprint [0]{\href }%
\providecommand \doibase [0]{https://doi.org/}%
\providecommand \selectlanguage [0]{\@gobble}%
\providecommand \bibinfo  [0]{\@secondoftwo}%
\providecommand \bibfield  [0]{\@secondoftwo}%
\providecommand \translation [1]{[#1]}%
\providecommand \BibitemOpen [0]{}%
\providecommand \bibitemStop [0]{}%
\providecommand \bibitemNoStop [0]{.\EOS\space}%
\providecommand \EOS [0]{\spacefactor3000\relax}%
\providecommand \BibitemShut  [1]{\csname bibitem#1\endcsname}%
\let\auto@bib@innerbib\@empty
\bibitem [{\citenamefont {Newnham}(2004)}]{newnham2004properties}%
  \BibitemOpen
  \bibfield  {author} {\bibinfo {author} {\bibfnamefont {R.~E.}\ \bibnamefont
  {Newnham}},\ }\href {https://doi.org/10.1093/oso/9780198520757.001.0001}
  {\emph {\bibinfo {title} {Properties of materials: anisotropy, symmetry,
  structure}}}\ (\bibinfo  {publisher} {OUP Oxford},\ \bibinfo {year}
  {2004})\BibitemShut {NoStop}%
\bibitem [{\citenamefont {Spaldin}\ and\ \citenamefont
  {Fiebig}(2005)}]{spaldin2005renaissance}%
  \BibitemOpen
  \bibfield  {author} {\bibinfo {author} {\bibfnamefont {N.~A.}\ \bibnamefont
  {Spaldin}}\ and\ \bibinfo {author} {\bibfnamefont {M.}~\bibnamefont
  {Fiebig}},\ }\bibfield  {title} {\bibinfo {title} {The renaissance of
  magnetoelectric multiferroics},\ }\href
  {https://doi.org/10.1126/science.1113357} {\bibfield  {journal} {\bibinfo
  {journal} {Science}\ }\textbf {\bibinfo {volume} {309}},\ \bibinfo {pages}
  {391} (\bibinfo {year} {2005})}\BibitemShut {NoStop}%
\bibitem [{\citenamefont {Spaldin}\ and\ \citenamefont
  {Ramesh}(2019)}]{spaldin2019advances}%
  \BibitemOpen
  \bibfield  {author} {\bibinfo {author} {\bibfnamefont {N.~A.}\ \bibnamefont
  {Spaldin}}\ and\ \bibinfo {author} {\bibfnamefont {R.}~\bibnamefont
  {Ramesh}},\ }\bibfield  {title} {\bibinfo {title} {Advances in
  magnetoelectric multiferroics},\ }\href
  {https://doi.org/10.1038/s41563-018-0275-2} {\bibfield  {journal} {\bibinfo
  {journal} {Nature materials}\ }\textbf {\bibinfo {volume} {18}},\ \bibinfo
  {pages} {203} (\bibinfo {year} {2019})}\BibitemShut {NoStop}%
\bibitem [{\citenamefont {Liang}\ \emph {et~al.}(2021)\citenamefont {Liang},
  \citenamefont {Matyushov}, \citenamefont {Hayes}, \citenamefont {Schell},
  \citenamefont {Dong}, \citenamefont {Chen}, \citenamefont {He}, \citenamefont
  {Will-Cole}, \citenamefont {Quandt}, \citenamefont {Martins} \emph
  {et~al.}}]{liang2021roadmap}%
  \BibitemOpen
  \bibfield  {author} {\bibinfo {author} {\bibfnamefont {X.}~\bibnamefont
  {Liang}}, \bibinfo {author} {\bibfnamefont {A.}~\bibnamefont {Matyushov}},
  \bibinfo {author} {\bibfnamefont {P.}~\bibnamefont {Hayes}}, \bibinfo
  {author} {\bibfnamefont {V.}~\bibnamefont {Schell}}, \bibinfo {author}
  {\bibfnamefont {C.}~\bibnamefont {Dong}}, \bibinfo {author} {\bibfnamefont
  {H.}~\bibnamefont {Chen}}, \bibinfo {author} {\bibfnamefont {Y.}~\bibnamefont
  {He}}, \bibinfo {author} {\bibfnamefont {A.}~\bibnamefont {Will-Cole}},
  \bibinfo {author} {\bibfnamefont {E.}~\bibnamefont {Quandt}}, \bibinfo
  {author} {\bibfnamefont {P.}~\bibnamefont {Martins}}, \emph {et~al.},\
  }\bibfield  {title} {\bibinfo {title} {Roadmap on magnetoelectric materials
  and devices},\ }\href {https://doi.org/10.1109/TMAG.2021.3086635} {\bibfield
  {journal} {\bibinfo  {journal} {IEEE Transactions on Magnetics}\ }\textbf
  {\bibinfo {volume} {57}},\ \bibinfo {pages} {1} (\bibinfo {year}
  {2021})}\BibitemShut {NoStop}%
\bibitem [{\citenamefont {Krichevtsov}\ \emph {et~al.}(1996)\citenamefont
  {Krichevtsov}, \citenamefont {Pavlov}, \citenamefont {Pisarev},\ and\
  \citenamefont {Gridnev}}]{krichevtsov1996magnetoelectric}%
  \BibitemOpen
  \bibfield  {author} {\bibinfo {author} {\bibfnamefont {B.}~\bibnamefont
  {Krichevtsov}}, \bibinfo {author} {\bibfnamefont {V.}~\bibnamefont {Pavlov}},
  \bibinfo {author} {\bibfnamefont {R.}~\bibnamefont {Pisarev}},\ and\ \bibinfo
  {author} {\bibfnamefont {V.}~\bibnamefont {Gridnev}},\ }\bibfield  {title}
  {\bibinfo {title} {Magnetoelectric spectroscopy of electronic transitions in
  antiferromagnetic {Cr$_2$O$_3$}},\ }\href
  {https://doi.org/10.1103/PhysRevLett.76.4628} {\bibfield  {journal} {\bibinfo
   {journal} {Physical review letters}\ }\textbf {\bibinfo {volume} {76}},\
  \bibinfo {pages} {4628} (\bibinfo {year} {1996})}\BibitemShut {NoStop}%
\bibitem [{\citenamefont {Bilyk}\ \emph {et~al.}(2025)\citenamefont {Bilyk},
  \citenamefont {Dubrovin}, \citenamefont {Zvezdin}, \citenamefont {Kirilyuk},\
  and\ \citenamefont {Kimel}}]{bilyk2025control}%
  \BibitemOpen
  \bibfield  {author} {\bibinfo {author} {\bibfnamefont {V.~R.}\ \bibnamefont
  {Bilyk}}, \bibinfo {author} {\bibfnamefont {R.~M.}\ \bibnamefont {Dubrovin}},
  \bibinfo {author} {\bibfnamefont {A.~K.}\ \bibnamefont {Zvezdin}}, \bibinfo
  {author} {\bibfnamefont {A.~I.}\ \bibnamefont {Kirilyuk}},\ and\ \bibinfo
  {author} {\bibfnamefont {A.~V.}\ \bibnamefont {Kimel}},\ }\bibfield  {title}
  {\bibinfo {title} {Control of spins in collinear antiferromagnet
  {Cr$_2$O$_3$} by terahertz electric fields},\ }\bibfield  {journal} {\bibinfo
   {journal} {Newton}\ }\href {https://doi.org/10.1016/j.newton.2025.100132}
  {10.1016/j.newton.2025.100132} (\bibinfo {year} {2025})\BibitemShut {NoStop}%
\bibitem [{\citenamefont {He}\ \emph {et~al.}(2025)\citenamefont {He},
  \citenamefont {Liu}, \citenamefont {Su}, \citenamefont {Hong},\ and\
  \citenamefont {Sun}}]{he2025dynamic}%
  \BibitemOpen
  \bibfield  {author} {\bibinfo {author} {\bibfnamefont {Y.}~\bibnamefont
  {He}}, \bibinfo {author} {\bibfnamefont {Z.}~\bibnamefont {Liu}}, \bibinfo
  {author} {\bibfnamefont {N.}~\bibnamefont {Su}}, \bibinfo {author}
  {\bibfnamefont {D.}~\bibnamefont {Hong}},\ and\ \bibinfo {author}
  {\bibfnamefont {Y.}~\bibnamefont {Sun}},\ }\bibfield  {title} {\bibinfo
  {title} {Dynamic magnetoelectric effect and phase diagram of the polar magnet
  {Ni$_2$MnTeO$_6$}},\ }\href {https://doi.org/10.1103/znqk-sfvb} {\bibfield
  {journal} {\bibinfo  {journal} {Physical Review B}\ }\textbf {\bibinfo
  {volume} {112}},\ \bibinfo {pages} {054422} (\bibinfo {year}
  {2025})}\BibitemShut {NoStop}%
\bibitem [{\citenamefont {Malashevich}\ \emph {et~al.}(2012)\citenamefont
  {Malashevich}, \citenamefont {Coh}, \citenamefont {Souza},\ and\
  \citenamefont {Vanderbilt}}]{malashevich2012full}%
  \BibitemOpen
  \bibfield  {author} {\bibinfo {author} {\bibfnamefont {A.}~\bibnamefont
  {Malashevich}}, \bibinfo {author} {\bibfnamefont {S.}~\bibnamefont {Coh}},
  \bibinfo {author} {\bibfnamefont {I.}~\bibnamefont {Souza}},\ and\ \bibinfo
  {author} {\bibfnamefont {D.}~\bibnamefont {Vanderbilt}},\ }\bibfield  {title}
  {\bibinfo {title} {Full magnetoelectric response of {Cr$_2$O$_3$} from first
  principles},\ }\href {https://doi.org/10.1103/PhysRevB.86.094430} {\bibfield
  {journal} {\bibinfo  {journal} {Physical Review B—Condensed Matter and
  Materials Physics}\ }\textbf {\bibinfo {volume} {86}},\ \bibinfo {pages}
  {094430} (\bibinfo {year} {2012})}\BibitemShut {NoStop}%
\bibitem [{\citenamefont {Malashevich}\ \emph {et~al.}(2010)\citenamefont
  {Malashevich}, \citenamefont {Souza}, \citenamefont {Coh},\ and\
  \citenamefont {Vanderbilt}}]{malashevich2010theory}%
  \BibitemOpen
  \bibfield  {author} {\bibinfo {author} {\bibfnamefont {A.}~\bibnamefont
  {Malashevich}}, \bibinfo {author} {\bibfnamefont {I.}~\bibnamefont {Souza}},
  \bibinfo {author} {\bibfnamefont {S.}~\bibnamefont {Coh}},\ and\ \bibinfo
  {author} {\bibfnamefont {D.}~\bibnamefont {Vanderbilt}},\ }\bibfield  {title}
  {\bibinfo {title} {Theory of orbital magnetoelectric response},\ }\href
  {https://doi.org/10.1088/1367-2630/12/5/053032} {\bibfield  {journal}
  {\bibinfo  {journal} {New Journal of Physics}\ }\textbf {\bibinfo {volume}
  {12}},\ \bibinfo {pages} {053032} (\bibinfo {year} {2010})}\BibitemShut
  {NoStop}%
\bibitem [{\citenamefont {Ye}\ and\ \citenamefont
  {Vanderbilt}(2014)}]{ye2014dynamical}%
  \BibitemOpen
  \bibfield  {author} {\bibinfo {author} {\bibfnamefont {M.}~\bibnamefont
  {Ye}}\ and\ \bibinfo {author} {\bibfnamefont {D.}~\bibnamefont
  {Vanderbilt}},\ }\bibfield  {title} {\bibinfo {title} {Dynamical magnetic
  charges and linear magnetoelectricity},\ }\href
  {https://doi.org/10.1103/PhysRevB.89.064301} {\bibfield  {journal} {\bibinfo
  {journal} {Physical Review B}\ }\textbf {\bibinfo {volume} {89}},\ \bibinfo
  {pages} {064301} (\bibinfo {year} {2014})}\BibitemShut {NoStop}%
\bibitem [{\citenamefont {Vanderbilt}(2018)}]{vanderbilt2018berry}%
  \BibitemOpen
  \bibfield  {author} {\bibinfo {author} {\bibfnamefont {D.}~\bibnamefont
  {Vanderbilt}},\ }\href {https://doi.org/10.1017/9781316662205} {\emph
  {\bibinfo {title} {Berry phases in electronic structure theory: electric
  polarization, orbital magnetization and topological insulators}}}\ (\bibinfo
  {publisher} {Cambridge University Press},\ \bibinfo {year}
  {2018})\BibitemShut {NoStop}%
\bibitem [{\citenamefont {Essin}\ \emph {et~al.}(2010)\citenamefont {Essin},
  \citenamefont {Turner}, \citenamefont {Moore},\ and\ \citenamefont
  {Vanderbilt}}]{essin2010orbital}%
  \BibitemOpen
  \bibfield  {author} {\bibinfo {author} {\bibfnamefont {A.~M.}\ \bibnamefont
  {Essin}}, \bibinfo {author} {\bibfnamefont {A.~M.}\ \bibnamefont {Turner}},
  \bibinfo {author} {\bibfnamefont {J.~E.}\ \bibnamefont {Moore}},\ and\
  \bibinfo {author} {\bibfnamefont {D.}~\bibnamefont {Vanderbilt}},\ }\bibfield
   {title} {\bibinfo {title} {Orbital magnetoelectric coupling in band
  insulators},\ }\href {https://doi.org/10.1103/PhysRevB.81.205104} {\bibfield
  {journal} {\bibinfo  {journal} {Physical Review B—Condensed Matter and
  Materials Physics}\ }\textbf {\bibinfo {volume} {81}},\ \bibinfo {pages}
  {205104} (\bibinfo {year} {2010})}\BibitemShut {NoStop}%
\bibitem [{\citenamefont {Brown~Jr}\ \emph {et~al.}(1968)\citenamefont
  {Brown~Jr}, \citenamefont {Hornreich},\ and\ \citenamefont
  {Shtrikman}}]{brown1968upper}%
  \BibitemOpen
  \bibfield  {author} {\bibinfo {author} {\bibfnamefont {W.}~\bibnamefont
  {Brown~Jr}}, \bibinfo {author} {\bibfnamefont {R.}~\bibnamefont
  {Hornreich}},\ and\ \bibinfo {author} {\bibfnamefont {S.}~\bibnamefont
  {Shtrikman}},\ }\bibfield  {title} {\bibinfo {title} {Upper bound on the
  magnetoelectric susceptibility},\ }\href
  {https://doi.org/10.1103/PhysRev.168.574} {\bibfield  {journal} {\bibinfo
  {journal} {Physical Review}\ }\textbf {\bibinfo {volume} {168}},\ \bibinfo
  {pages} {574} (\bibinfo {year} {1968})}\BibitemShut {NoStop}%
\bibitem [{\citenamefont {Fiebig}(2005)}]{fiebig2004revival}%
  \BibitemOpen
  \bibfield  {author} {\bibinfo {author} {\bibfnamefont {M.}~\bibnamefont
  {Fiebig}},\ }\bibfield  {title} {\bibinfo {title} {Revival of the
  magnetoelectric effect},\ }\href {https://doi.org/10.1088/0022-3727/38/8/R01}
  {\bibfield  {journal} {\bibinfo  {journal} {Journal of physics D: applied
  physics}\ }\textbf {\bibinfo {volume} {38}},\ \bibinfo {pages} {R123}
  (\bibinfo {year} {2005})}\BibitemShut {NoStop}%
\bibitem [{\citenamefont {Vopson}\ \emph {et~al.}(2017)\citenamefont {Vopson},
  \citenamefont {Fetisov}, \citenamefont {Caruntu},\ and\ \citenamefont
  {Srinivasan}}]{vopson2017measurement}%
  \BibitemOpen
  \bibfield  {author} {\bibinfo {author} {\bibfnamefont {M.}~\bibnamefont
  {Vopson}}, \bibinfo {author} {\bibfnamefont {Y.}~\bibnamefont {Fetisov}},
  \bibinfo {author} {\bibfnamefont {G.}~\bibnamefont {Caruntu}},\ and\ \bibinfo
  {author} {\bibfnamefont {G.}~\bibnamefont {Srinivasan}},\ }\bibfield  {title}
  {\bibinfo {title} {Measurement techniques of the magneto-electric coupling in
  multiferroics},\ }\href {https://doi.org/10.3390/ma10080963} {\bibfield
  {journal} {\bibinfo  {journal} {materials}\ }\textbf {\bibinfo {volume}
  {10}},\ \bibinfo {pages} {963} (\bibinfo {year} {2017})}\BibitemShut
  {NoStop}%
\bibitem [{\citenamefont {Y{\=\i}ng}\ and\ \citenamefont
  {Z{\"u}licke}(2022)}]{ying2022magnetoelectricity}%
  \BibitemOpen
  \bibfield  {author} {\bibinfo {author} {\bibfnamefont {Y.}~\bibnamefont
  {Y{\=\i}ng}}\ and\ \bibinfo {author} {\bibfnamefont {U.}~\bibnamefont
  {Z{\"u}licke}},\ }\bibfield  {title} {\bibinfo {title} {Magnetoelectricity in
  two-dimensional materials},\ }\href
  {https://doi.org/10.1080/23746149.2022.2032343} {\bibfield  {journal}
  {\bibinfo  {journal} {Advances in Physics: X}\ }\textbf {\bibinfo {volume}
  {7}},\ \bibinfo {pages} {2032343} (\bibinfo {year} {2022})}\BibitemShut
  {NoStop}%
\bibitem [{\citenamefont {Bousquet}\ \emph {et~al.}(2011)\citenamefont
  {Bousquet}, \citenamefont {Spaldin},\ and\ \citenamefont
  {Delaney}}]{bousquet2011unexpectedly}%
  \BibitemOpen
  \bibfield  {author} {\bibinfo {author} {\bibfnamefont {E.}~\bibnamefont
  {Bousquet}}, \bibinfo {author} {\bibfnamefont {N.~A.}\ \bibnamefont
  {Spaldin}},\ and\ \bibinfo {author} {\bibfnamefont {K.~T.}\ \bibnamefont
  {Delaney}},\ }\bibfield  {title} {\bibinfo {title} {Unexpectedly large
  electronic contribution to linear magnetoelectricity},\ }\href
  {https://doi.org/10.1103/PhysRevLett.106.107202} {\bibfield  {journal}
  {\bibinfo  {journal} {Physical Review Letters}\ }\textbf {\bibinfo {volume}
  {106}},\ \bibinfo {pages} {107202} (\bibinfo {year} {2011})}\BibitemShut
  {NoStop}%
\bibitem [{\citenamefont {Hayashida}\ \emph {et~al.}(2022)\citenamefont
  {Hayashida}, \citenamefont {Arakawa}, \citenamefont {Oshima}, \citenamefont
  {Kimura},\ and\ \citenamefont {Kimura}}]{hayashida2022observation}%
  \BibitemOpen
  \bibfield  {author} {\bibinfo {author} {\bibfnamefont {T.}~\bibnamefont
  {Hayashida}}, \bibinfo {author} {\bibfnamefont {K.}~\bibnamefont {Arakawa}},
  \bibinfo {author} {\bibfnamefont {T.}~\bibnamefont {Oshima}}, \bibinfo
  {author} {\bibfnamefont {K.}~\bibnamefont {Kimura}},\ and\ \bibinfo {author}
  {\bibfnamefont {T.}~\bibnamefont {Kimura}},\ }\bibfield  {title} {\bibinfo
  {title} {Observation of antiferromagnetic domains in {Cr$_2$O$_3$} using
  nonreciprocal optical effects},\ }\href
  {https://doi.org/10.1103/PhysRevResearch.4.043063} {\bibfield  {journal}
  {\bibinfo  {journal} {Physical Review Research}\ }\textbf {\bibinfo {volume}
  {4}},\ \bibinfo {pages} {043063} (\bibinfo {year} {2022})}\BibitemShut
  {NoStop}%
\bibitem [{\citenamefont {Muto}\ \emph {et~al.}(1998)\citenamefont {Muto},
  \citenamefont {Tanabe}, \citenamefont {Iizuka-Sakano},\ and\ \citenamefont
  {Hanamura}}]{muto1998magnetoelectric}%
  \BibitemOpen
  \bibfield  {author} {\bibinfo {author} {\bibfnamefont {M.}~\bibnamefont
  {Muto}}, \bibinfo {author} {\bibfnamefont {Y.}~\bibnamefont {Tanabe}},
  \bibinfo {author} {\bibfnamefont {T.}~\bibnamefont {Iizuka-Sakano}},\ and\
  \bibinfo {author} {\bibfnamefont {E.}~\bibnamefont {Hanamura}},\ }\bibfield
  {title} {\bibinfo {title} {Magnetoelectric and second-harmonic spectra in
  antiferromagnetic {Cr$_2$O$_3$}},\ }\href
  {https://doi.org/10.1103/PhysRevB.57.9586} {\bibfield  {journal} {\bibinfo
  {journal} {Physical Review B}\ }\textbf {\bibinfo {volume} {57}},\ \bibinfo
  {pages} {9586} (\bibinfo {year} {1998})}\BibitemShut {NoStop}%
\bibitem [{\citenamefont {Kaxiras}\ and\ \citenamefont
  {Joannopoulos}(2019)}]{kaxiras2019quantum}%
  \BibitemOpen
  \bibfield  {author} {\bibinfo {author} {\bibfnamefont {E.}~\bibnamefont
  {Kaxiras}}\ and\ \bibinfo {author} {\bibfnamefont {J.~D.}\ \bibnamefont
  {Joannopoulos}},\ }\href {https://doi.org/10.1017/9781139030809} {\emph
  {\bibinfo {title} {Quantum theory of materials}}}\ (\bibinfo  {publisher}
  {Cambridge university press},\ \bibinfo {year} {2019})\BibitemShut {NoStop}%
\bibitem [{\citenamefont {Rzazewski}\ and\ \citenamefont
  {Boyd}(2004)}]{rzkazewski2004equivalence}%
  \BibitemOpen
  \bibfield  {author} {\bibinfo {author} {\bibfnamefont {K.}~\bibnamefont
  {Rzazewski}}\ and\ \bibinfo {author} {\bibfnamefont {R.~W.}\ \bibnamefont
  {Boyd}},\ }\bibfield  {title} {\bibinfo {title} {Equivalence of interaction
  hamiltonians in the electric dipole approximation},\ }\href
  {https://doi.org/10.1080/09500340408230412} {\bibfield  {journal} {\bibinfo
  {journal} {Journal of modern optics}\ }\textbf {\bibinfo {volume} {51}},\
  \bibinfo {pages} {1137} (\bibinfo {year} {2004})}\BibitemShut {NoStop}%
\bibitem [{\citenamefont {Pozo~Oca{\~n}a}\ and\ \citenamefont
  {Souza}(2023)}]{pozo2023multipole}%
  \BibitemOpen
  \bibfield  {author} {\bibinfo {author} {\bibfnamefont {{\'O}.}~\bibnamefont
  {Pozo~Oca{\~n}a}}\ and\ \bibinfo {author} {\bibfnamefont {I.}~\bibnamefont
  {Souza}},\ }\bibfield  {title} {\bibinfo {title} {Multipole theory of optical
  spatial dispersion in crystals},\ }\href
  {https://doi.org/10.21468/SciPostPhys.14.5.118} {\bibfield  {journal}
  {\bibinfo  {journal} {SciPost Physics}\ }\textbf {\bibinfo {volume} {14}},\
  \bibinfo {pages} {118} (\bibinfo {year} {2023})}\BibitemShut {NoStop}%
\bibitem [{\citenamefont {Marsili}\ \emph {et~al.}(2021)\citenamefont
  {Marsili}, \citenamefont {Molina-S{\'a}nchez}, \citenamefont {Palummo},
  \citenamefont {Sangalli},\ and\ \citenamefont
  {Marini}}]{marsili2021spinorial}%
  \BibitemOpen
  \bibfield  {author} {\bibinfo {author} {\bibfnamefont {M.}~\bibnamefont
  {Marsili}}, \bibinfo {author} {\bibfnamefont {A.}~\bibnamefont
  {Molina-S{\'a}nchez}}, \bibinfo {author} {\bibfnamefont {M.}~\bibnamefont
  {Palummo}}, \bibinfo {author} {\bibfnamefont {D.}~\bibnamefont {Sangalli}},\
  and\ \bibinfo {author} {\bibfnamefont {A.}~\bibnamefont {Marini}},\
  }\bibfield  {title} {\bibinfo {title} {Spinorial formulation of the
  {$GW$-BSE} equations and spin properties of excitons in two-dimensional
  transition metal dichalcogenides},\ }\href
  {https://doi.org/10.1103/PhysRevB.103.155152} {\bibfield  {journal} {\bibinfo
   {journal} {Physical Review B}\ }\textbf {\bibinfo {volume} {103}},\ \bibinfo
  {pages} {155152} (\bibinfo {year} {2021})}\BibitemShut {NoStop}%
\bibitem [{\citenamefont {Wu}\ \emph {et~al.}(2022)\citenamefont {Wu},
  \citenamefont {Li},\ and\ \citenamefont {Louie}}]{wu2022optical}%
  \BibitemOpen
  \bibfield  {author} {\bibinfo {author} {\bibfnamefont {M.}~\bibnamefont
  {Wu}}, \bibinfo {author} {\bibfnamefont {Z.}~\bibnamefont {Li}},\ and\
  \bibinfo {author} {\bibfnamefont {S.~G.}\ \bibnamefont {Louie}},\ }\bibfield
  {title} {\bibinfo {title} {Optical and magneto-optical properties of
  ferromagnetic monolayer {CrBr$_3$}: A first-principles {$GW$} and {$GW$} plus
  {Bethe--Salpeter} equation study},\ }\href
  {https://doi.org/10.1103/PhysRevMaterials.6.014008} {\bibfield  {journal}
  {\bibinfo  {journal} {Physical Review Materials}\ }\textbf {\bibinfo {volume}
  {6}},\ \bibinfo {pages} {014008} (\bibinfo {year} {2022})}\BibitemShut
  {NoStop}%
\bibitem [{\citenamefont {Martin}\ \emph {et~al.}(2016)\citenamefont {Martin},
  \citenamefont {Reining},\ and\ \citenamefont
  {Ceperley}}]{martin2016interacting}%
  \BibitemOpen
  \bibfield  {author} {\bibinfo {author} {\bibfnamefont {R.~M.}\ \bibnamefont
  {Martin}}, \bibinfo {author} {\bibfnamefont {L.}~\bibnamefont {Reining}},\
  and\ \bibinfo {author} {\bibfnamefont {D.~M.}\ \bibnamefont {Ceperley}},\
  }\href {https://doi.org/10.1017/CBO9781139050807} {\emph {\bibinfo {title}
  {Interacting electrons}}}\ (\bibinfo  {publisher} {Cambridge University
  Press},\ \bibinfo {year} {2016})\BibitemShut {NoStop}%
\bibitem [{\citenamefont {Reining}\ \emph {et~al.}(2002)\citenamefont
  {Reining}, \citenamefont {Olevano}, \citenamefont {Rubio},\ and\
  \citenamefont {Onida}}]{reining2002excitonic}%
  \BibitemOpen
  \bibfield  {author} {\bibinfo {author} {\bibfnamefont {L.}~\bibnamefont
  {Reining}}, \bibinfo {author} {\bibfnamefont {V.}~\bibnamefont {Olevano}},
  \bibinfo {author} {\bibfnamefont {A.}~\bibnamefont {Rubio}},\ and\ \bibinfo
  {author} {\bibfnamefont {G.}~\bibnamefont {Onida}},\ }\bibfield  {title}
  {\bibinfo {title} {Excitonic effects in solids described by time-dependent
  density-functional theory},\ }\href
  {https://doi.org/10.1103/PhysRevLett.88.066404} {\bibfield  {journal}
  {\bibinfo  {journal} {Physical review letters}\ }\textbf {\bibinfo {volume}
  {88}},\ \bibinfo {pages} {066404} (\bibinfo {year} {2002})}\BibitemShut
  {NoStop}%
\bibitem [{\citenamefont {Guandalini}\ \emph {et~al.}(2023)\citenamefont
  {Guandalini}, \citenamefont {D’Amico}, \citenamefont {Ferretti},\ and\
  \citenamefont {Varsano}}]{guandalini2023efficient}%
  \BibitemOpen
  \bibfield  {author} {\bibinfo {author} {\bibfnamefont {A.}~\bibnamefont
  {Guandalini}}, \bibinfo {author} {\bibfnamefont {P.}~\bibnamefont
  {D’Amico}}, \bibinfo {author} {\bibfnamefont {A.}~\bibnamefont
  {Ferretti}},\ and\ \bibinfo {author} {\bibfnamefont {D.}~\bibnamefont
  {Varsano}},\ }\bibfield  {title} {\bibinfo {title} {Efficient {$GW$}
  calculations in two dimensional materials through a stochastic integration of
  the screened potential},\ }\href {https://doi.org/10.1038/s41524-023-00989-7}
  {\bibfield  {journal} {\bibinfo  {journal} {npj Computational Materials}\
  }\textbf {\bibinfo {volume} {9}},\ \bibinfo {pages} {44} (\bibinfo {year}
  {2023})}\BibitemShut {NoStop}%
\bibitem [{Note1()}]{Note1}%
  \BibitemOpen
  \bibinfo {note} {For both the LDA and the GGA ground state, in the \protect
  \textsc {Lumen} code, only the term containing the functional derivative with
  respect to the density is constructed in the definition of the adiabatic
  kernel, i.e. $f^A_{xc}=\partial v^{DFT}[\rho _{\alpha \beta }(t)](\protect
  \mathbf {x},t)/\partial \rho _{\alpha \beta }(\protect \mathbf
  {x'},t)$.}\BibitemShut {Stop}%
\bibitem [{\citenamefont {Giannozzi}\ \emph {et~al.}(2017)\citenamefont
  {Giannozzi}, \citenamefont {Andreussi}, \citenamefont {Brumme}, \citenamefont
  {Bunau}, \citenamefont {Nardelli}, \citenamefont {Calandra}, \citenamefont
  {Car}, \citenamefont {Cavazzoni}, \citenamefont {Ceresoli}, \citenamefont
  {Cococcioni}, \citenamefont {Colonna}, \citenamefont {Carnimeo},
  \citenamefont {Corso}, \citenamefont {de~Gironcoli}, \citenamefont {Delugas},
  \citenamefont {Jr}, \citenamefont {Ferretti}, \citenamefont {Floris},
  \citenamefont {Fratesi}, \citenamefont {Fugallo}, \citenamefont {Gebauer},
  \citenamefont {Gerstmann}, \citenamefont {Giustino}, \citenamefont {Gorni},
  \citenamefont {Jia}, \citenamefont {Kawamura}, \citenamefont {Ko},
  \citenamefont {Kokalj}, \citenamefont {Küçükbenli}, \citenamefont
  {Lazzeri}, \citenamefont {Marsili}, \citenamefont {Marzari}, \citenamefont
  {Mauri}, \citenamefont {Nguyen}, \citenamefont {Nguyen}, \citenamefont {de-la
  Roza}, \citenamefont {Paulatto}, \citenamefont {Poncé}, \citenamefont
  {Rocca}, \citenamefont {Sabatini}, \citenamefont {Santra}, \citenamefont
  {Schlipf}, \citenamefont {Seitsonen}, \citenamefont {Smogunov}, \citenamefont
  {Timrov}, \citenamefont {Thonhauser}, \citenamefont {Umari}, \citenamefont
  {Vast}, \citenamefont {Wu},\ and\ \citenamefont {Baroni}}]{QE-2017}%
  \BibitemOpen
  \bibfield  {author} {\bibinfo {author} {\bibfnamefont {P.}~\bibnamefont
  {Giannozzi}}, \bibinfo {author} {\bibfnamefont {O.}~\bibnamefont
  {Andreussi}}, \bibinfo {author} {\bibfnamefont {T.}~\bibnamefont {Brumme}},
  \bibinfo {author} {\bibfnamefont {O.}~\bibnamefont {Bunau}}, \bibinfo
  {author} {\bibfnamefont {M.~B.}\ \bibnamefont {Nardelli}}, \bibinfo {author}
  {\bibfnamefont {M.}~\bibnamefont {Calandra}}, \bibinfo {author}
  {\bibfnamefont {R.}~\bibnamefont {Car}}, \bibinfo {author} {\bibfnamefont
  {C.}~\bibnamefont {Cavazzoni}}, \bibinfo {author} {\bibfnamefont
  {D.}~\bibnamefont {Ceresoli}}, \bibinfo {author} {\bibfnamefont
  {M.}~\bibnamefont {Cococcioni}}, \bibinfo {author} {\bibfnamefont
  {N.}~\bibnamefont {Colonna}}, \bibinfo {author} {\bibfnamefont
  {I.}~\bibnamefont {Carnimeo}}, \bibinfo {author} {\bibfnamefont {A.~D.}\
  \bibnamefont {Corso}}, \bibinfo {author} {\bibfnamefont {S.}~\bibnamefont
  {de~Gironcoli}}, \bibinfo {author} {\bibfnamefont {P.}~\bibnamefont
  {Delugas}}, \bibinfo {author} {\bibfnamefont {R.~A.~D.}\ \bibnamefont {Jr}},
  \bibinfo {author} {\bibfnamefont {A.}~\bibnamefont {Ferretti}}, \bibinfo
  {author} {\bibfnamefont {A.}~\bibnamefont {Floris}}, \bibinfo {author}
  {\bibfnamefont {G.}~\bibnamefont {Fratesi}}, \bibinfo {author} {\bibfnamefont
  {G.}~\bibnamefont {Fugallo}}, \bibinfo {author} {\bibfnamefont
  {R.}~\bibnamefont {Gebauer}}, \bibinfo {author} {\bibfnamefont
  {U.}~\bibnamefont {Gerstmann}}, \bibinfo {author} {\bibfnamefont
  {F.}~\bibnamefont {Giustino}}, \bibinfo {author} {\bibfnamefont
  {T.}~\bibnamefont {Gorni}}, \bibinfo {author} {\bibfnamefont
  {J.}~\bibnamefont {Jia}}, \bibinfo {author} {\bibfnamefont {M.}~\bibnamefont
  {Kawamura}}, \bibinfo {author} {\bibfnamefont {H.-Y.}\ \bibnamefont {Ko}},
  \bibinfo {author} {\bibfnamefont {A.}~\bibnamefont {Kokalj}}, \bibinfo
  {author} {\bibfnamefont {E.}~\bibnamefont {Küçükbenli}}, \bibinfo {author}
  {\bibfnamefont {M.}~\bibnamefont {Lazzeri}}, \bibinfo {author} {\bibfnamefont
  {M.}~\bibnamefont {Marsili}}, \bibinfo {author} {\bibfnamefont
  {N.}~\bibnamefont {Marzari}}, \bibinfo {author} {\bibfnamefont
  {F.}~\bibnamefont {Mauri}}, \bibinfo {author} {\bibfnamefont {N.~L.}\
  \bibnamefont {Nguyen}}, \bibinfo {author} {\bibfnamefont {H.-V.}\
  \bibnamefont {Nguyen}}, \bibinfo {author} {\bibfnamefont {A.~O.}\
  \bibnamefont {de-la Roza}}, \bibinfo {author} {\bibfnamefont
  {L.}~\bibnamefont {Paulatto}}, \bibinfo {author} {\bibfnamefont
  {S.}~\bibnamefont {Poncé}}, \bibinfo {author} {\bibfnamefont
  {D.}~\bibnamefont {Rocca}}, \bibinfo {author} {\bibfnamefont
  {R.}~\bibnamefont {Sabatini}}, \bibinfo {author} {\bibfnamefont
  {B.}~\bibnamefont {Santra}}, \bibinfo {author} {\bibfnamefont
  {M.}~\bibnamefont {Schlipf}}, \bibinfo {author} {\bibfnamefont {A.~P.}\
  \bibnamefont {Seitsonen}}, \bibinfo {author} {\bibfnamefont {A.}~\bibnamefont
  {Smogunov}}, \bibinfo {author} {\bibfnamefont {I.}~\bibnamefont {Timrov}},
  \bibinfo {author} {\bibfnamefont {T.}~\bibnamefont {Thonhauser}}, \bibinfo
  {author} {\bibfnamefont {P.}~\bibnamefont {Umari}}, \bibinfo {author}
  {\bibfnamefont {N.}~\bibnamefont {Vast}}, \bibinfo {author} {\bibfnamefont
  {X.}~\bibnamefont {Wu}},\ and\ \bibinfo {author} {\bibfnamefont
  {S.}~\bibnamefont {Baroni}},\ }\bibfield  {title} {\bibinfo {title} {Advanced
  capabilities for materials modelling with {QUANTUM ESPRESSO}},\ }\href
  {https://doi.org/10.1088/1361-648X/aa8f79} {\bibfield  {journal} {\bibinfo
  {journal} {Journal of Physics: Condensed Matter}\ }\textbf {\bibinfo {volume}
  {29}},\ \bibinfo {pages} {465901} (\bibinfo {year} {2017})}\BibitemShut
  {NoStop}%
\bibitem [{\citenamefont {Perdew}\ \emph {et~al.}(1996)\citenamefont {Perdew},
  \citenamefont {Burke},\ and\ \citenamefont {Ernzerhof}}]{PBE1996}%
  \BibitemOpen
  \bibfield  {author} {\bibinfo {author} {\bibfnamefont {J.~P.}\ \bibnamefont
  {Perdew}}, \bibinfo {author} {\bibfnamefont {K.}~\bibnamefont {Burke}},\ and\
  \bibinfo {author} {\bibfnamefont {M.}~\bibnamefont {Ernzerhof}},\ }\bibfield
  {title} {\bibinfo {title} {Generalized gradient approximation made simple},\
  }\href {https://doi.org/10.1103/PhysRevLett.77.3865} {\bibfield  {journal}
  {\bibinfo  {journal} {Physical Review Letters}\ }\textbf {\bibinfo {volume}
  {77}},\ \bibinfo {pages} {3865} (\bibinfo {year} {1996})}\BibitemShut
  {NoStop}%
\bibitem [{\citenamefont {Hamann}(2013)}]{Hamann2013_ONCVPSP}%
  \BibitemOpen
  \bibfield  {author} {\bibinfo {author} {\bibfnamefont {D.~R.}\ \bibnamefont
  {Hamann}},\ }\bibfield  {title} {\bibinfo {title} {Optimized norm-conserving
  vanderbilt pseudopotentials},\ }\href
  {https://doi.org/10.1103/PhysRevB.88.085117} {\bibfield  {journal} {\bibinfo
  {journal} {Physical Review B}\ }\textbf {\bibinfo {volume} {88}},\ \bibinfo
  {pages} {085117} (\bibinfo {year} {2013})}\BibitemShut {NoStop}%
\bibitem [{\citenamefont {Hill}\ \emph {et~al.}(2010)\citenamefont {Hill},
  \citenamefont {Harrison}, \citenamefont {Dickinson}, \citenamefont {Zhou},\
  and\ \citenamefont {Kockelmann}}]{hill2010crystallographic}%
  \BibitemOpen
  \bibfield  {author} {\bibinfo {author} {\bibfnamefont {A.~H.}\ \bibnamefont
  {Hill}}, \bibinfo {author} {\bibfnamefont {A.}~\bibnamefont {Harrison}},
  \bibinfo {author} {\bibfnamefont {C.}~\bibnamefont {Dickinson}}, \bibinfo
  {author} {\bibfnamefont {W.}~\bibnamefont {Zhou}},\ and\ \bibinfo {author}
  {\bibfnamefont {W.}~\bibnamefont {Kockelmann}},\ }\bibfield  {title}
  {\bibinfo {title} {Crystallographic and magnetic studies of mesoporous
  eskolaite, {Cr$_2$O$_3$}},\ }\href
  {https://doi.org/10.1016/j.micromeso.2009.11.021} {\bibfield  {journal}
  {\bibinfo  {journal} {Microporous and Mesoporous Materials}\ }\textbf
  {\bibinfo {volume} {130}},\ \bibinfo {pages} {280} (\bibinfo {year}
  {2010})}\BibitemShut {NoStop}%
\bibitem [{si()}]{si}%
  \BibitemOpen
  \href@noop {} {}\bibinfo {note} {"See Supplemental Material for the (i)
  geometry and (ii) convergence plots along with further comparisons between
  LDA and GGA, and (iii) the effect of DFT$+U$ on the BSE
  spectra."}\BibitemShut {NoStop}%
\bibitem [{\citenamefont {Perdew}\ and\ \citenamefont
  {Zunger}(1981)}]{PerdewZunger1981}%
  \BibitemOpen
  \bibfield  {author} {\bibinfo {author} {\bibfnamefont {J.~P.}\ \bibnamefont
  {Perdew}}\ and\ \bibinfo {author} {\bibfnamefont {A.}~\bibnamefont
  {Zunger}},\ }\bibfield  {title} {\bibinfo {title} {Self-interaction
  correction to density-functional approximations for many-electron systems},\
  }\href {https://doi.org/10.1103/PhysRevB.23.5048} {\bibfield  {journal}
  {\bibinfo  {journal} {Physical Review B}\ }\textbf {\bibinfo {volume} {23}},\
  \bibinfo {pages} {5048} (\bibinfo {year} {1981})}\BibitemShut {NoStop}%
\bibitem [{\citenamefont {Sangalli}\ \emph {et~al.}(2019)\citenamefont
  {Sangalli}, \citenamefont {Ferretti}, \citenamefont {Miranda}, \citenamefont
  {Attaccalite}, \citenamefont {Marri}, \citenamefont {Cannuccia},
  \citenamefont {Melo}, \citenamefont {Marsili}, \citenamefont {Paleari},
  \citenamefont {Marrazzo} \emph {et~al.}}]{sangalli2019many}%
  \BibitemOpen
  \bibfield  {author} {\bibinfo {author} {\bibfnamefont {D.}~\bibnamefont
  {Sangalli}}, \bibinfo {author} {\bibfnamefont {A.}~\bibnamefont {Ferretti}},
  \bibinfo {author} {\bibfnamefont {H.}~\bibnamefont {Miranda}}, \bibinfo
  {author} {\bibfnamefont {C.}~\bibnamefont {Attaccalite}}, \bibinfo {author}
  {\bibfnamefont {I.}~\bibnamefont {Marri}}, \bibinfo {author} {\bibfnamefont
  {E.}~\bibnamefont {Cannuccia}}, \bibinfo {author} {\bibfnamefont
  {P.}~\bibnamefont {Melo}}, \bibinfo {author} {\bibfnamefont {M.}~\bibnamefont
  {Marsili}}, \bibinfo {author} {\bibfnamefont {F.}~\bibnamefont {Paleari}},
  \bibinfo {author} {\bibfnamefont {A.}~\bibnamefont {Marrazzo}}, \emph
  {et~al.},\ }\bibfield  {title} {\bibinfo {title} {Many-body perturbation
  theory calculations using the yambo code},\ }\href
  {https://doi.org/10.1088/1361-648X/ab15d0} {\bibfield  {journal} {\bibinfo
  {journal} {Journal of Physics: Condensed Matter}\ }\textbf {\bibinfo {volume}
  {31}},\ \bibinfo {pages} {325902} (\bibinfo {year} {2019})}\BibitemShut
  {NoStop}%
\bibitem [{\citenamefont {Sangalli}\ \emph {et~al.}(2026)\citenamefont
  {Sangalli}, \citenamefont {Reho}, \citenamefont {Alliati}, \citenamefont
  {Cervantes-Villanueva}, \citenamefont {Esquembre~Kučukalić}, \citenamefont
  {Geirsson}, \citenamefont {Leon~Valido}, \citenamefont {Mellado~Pinto},
  \citenamefont {Milev}, \citenamefont {Nalabothula}, \citenamefont {Paleari},
  \citenamefont {Romani}, \citenamefont {Bellentani}, \citenamefont
  {Castelo~Ares}, \citenamefont {D'Alessandro}, \citenamefont {Laricchia},
  \citenamefont {Montagna}, \citenamefont {Orlandini}, \citenamefont {Roman},
  \citenamefont {Affinito}, \citenamefont {Molina-S\'anchez}, \citenamefont
  {Attaccalite},\ and\ \citenamefont {Gr\"uning}}]{lumen2025}%
  \BibitemOpen
  \bibfield  {author} {\bibinfo {author} {\bibfnamefont {D.}~\bibnamefont
  {Sangalli}}, \bibinfo {author} {\bibfnamefont {R.}~\bibnamefont {Reho}},
  \bibinfo {author} {\bibfnamefont {I.~M.}\ \bibnamefont {Alliati}}, \bibinfo
  {author} {\bibfnamefont {J.}~\bibnamefont {Cervantes-Villanueva}}, \bibinfo
  {author} {\bibfnamefont {A.}~\bibnamefont {Esquembre~Kučukalić}}, \bibinfo
  {author} {\bibfnamefont {T.}~\bibnamefont {Geirsson}}, \bibinfo {author}
  {\bibfnamefont {D.~A.}\ \bibnamefont {Leon~Valido}}, \bibinfo {author}
  {\bibfnamefont {B.}~\bibnamefont {Mellado~Pinto}}, \bibinfo {author}
  {\bibfnamefont {P.}~\bibnamefont {Milev}}, \bibinfo {author} {\bibfnamefont
  {M.}~\bibnamefont {Nalabothula}}, \bibinfo {author} {\bibfnamefont
  {F.}~\bibnamefont {Paleari}}, \bibinfo {author} {\bibfnamefont
  {A.}~\bibnamefont {Romani}}, \bibinfo {author} {\bibfnamefont
  {L.}~\bibnamefont {Bellentani}}, \bibinfo {author} {\bibfnamefont {J.~M.}\
  \bibnamefont {Castelo~Ares}}, \bibinfo {author} {\bibfnamefont
  {M.}~\bibnamefont {D'Alessandro}}, \bibinfo {author} {\bibfnamefont
  {S.}~\bibnamefont {Laricchia}}, \bibinfo {author} {\bibfnamefont {E.~M.}\
  \bibnamefont {Montagna}}, \bibinfo {author} {\bibfnamefont {S.}~\bibnamefont
  {Orlandini}}, \bibinfo {author} {\bibfnamefont {J.~E.}\ \bibnamefont
  {Roman}}, \bibinfo {author} {\bibfnamefont {F.}~\bibnamefont {Affinito}},
  \bibinfo {author} {\bibfnamefont {A.}~\bibnamefont {Molina-S\'anchez}},
  \bibinfo {author} {\bibfnamefont {C.}~\bibnamefont {Attaccalite}},\ and\
  \bibinfo {author} {\bibfnamefont {M.}~\bibnamefont {Gr\"uning}},\ }\href
  {https://doi.org/10.5281/zenodo.20121198} {\bibinfo {title} {Lumen}}
  (\bibinfo {year} {2026})\BibitemShut {NoStop}%
\bibitem [{\citenamefont {Alliati}\ \emph {et~al.}(2022)\citenamefont
  {Alliati}, \citenamefont {Sangalli},\ and\ \citenamefont
  {Gr{\"u}ning}}]{alliati2022double}%
  \BibitemOpen
  \bibfield  {author} {\bibinfo {author} {\bibfnamefont {I.~M.}\ \bibnamefont
  {Alliati}}, \bibinfo {author} {\bibfnamefont {D.}~\bibnamefont {Sangalli}},\
  and\ \bibinfo {author} {\bibfnamefont {M.}~\bibnamefont {Gr{\"u}ning}},\
  }\bibfield  {title} {\bibinfo {title} {Double k-grid method for solving the
  {Bethe--Salpeter} equation via {Lanczos} approaches},\ }\href
  {https://doi.org/10.3389/fchem.2021.763946} {\bibfield  {journal} {\bibinfo
  {journal} {Frontiers in Chemistry}\ }\textbf {\bibinfo {volume} {9}},\
  \bibinfo {pages} {763946} (\bibinfo {year} {2022})}\BibitemShut {NoStop}%
\bibitem [{\citenamefont {Zimmermann}\ \emph {et~al.}(1996)\citenamefont
  {Zimmermann}, \citenamefont {Steiner},\ and\ \citenamefont
  {H{\"u}fner}}]{zimmermann1996electron}%
  \BibitemOpen
  \bibfield  {author} {\bibinfo {author} {\bibfnamefont {R.}~\bibnamefont
  {Zimmermann}}, \bibinfo {author} {\bibfnamefont {P.}~\bibnamefont
  {Steiner}},\ and\ \bibinfo {author} {\bibfnamefont {S.}~\bibnamefont
  {H{\"u}fner}},\ }\bibfield  {title} {\bibinfo {title} {Electron
  spectroscopies and partial excitation spectra in {Cr$_2$O$_3$}},\ }in\ \href
  {https://doi.org/10.1016/S0368-2048(96)80024-7} {\emph {\bibinfo {booktitle}
  {Proceedings of the 11th International Conference on Vacuum Ultraviolet
  Radiation Physics}}}\ (\bibinfo {organization} {Elsevier},\ \bibinfo {year}
  {1996})\ pp.\ \bibinfo {pages} {49--52}\BibitemShut {NoStop}%
\bibitem [{\citenamefont {Lebreau}\ \emph {et~al.}(2014)\citenamefont
  {Lebreau}, \citenamefont {Islam}, \citenamefont {Diawara},\ and\
  \citenamefont {Marcus}}]{lebreau2014structural}%
  \BibitemOpen
  \bibfield  {author} {\bibinfo {author} {\bibfnamefont {F.}~\bibnamefont
  {Lebreau}}, \bibinfo {author} {\bibfnamefont {M.~M.}\ \bibnamefont {Islam}},
  \bibinfo {author} {\bibfnamefont {B.}~\bibnamefont {Diawara}},\ and\ \bibinfo
  {author} {\bibfnamefont {P.}~\bibnamefont {Marcus}},\ }\bibfield  {title}
  {\bibinfo {title} {Structural, magnetic, electronic, defect, and diffusion
  properties of {Cr$_2$O$_3$}: A {DFT+$U$} study},\ }\href
  {https://doi.org/10.1021/jp5039943} {\bibfield  {journal} {\bibinfo
  {journal} {The Journal of Physical Chemistry C}\ }\textbf {\bibinfo {volume}
  {118}},\ \bibinfo {pages} {18133} (\bibinfo {year} {2014})}\BibitemShut
  {NoStop}%
\bibitem [{\citenamefont {Shi}\ \emph {et~al.}(2009)\citenamefont {Shi},
  \citenamefont {Wysocki},\ and\ \citenamefont
  {Belashchenko}}]{shi2009magnetism}%
  \BibitemOpen
  \bibfield  {author} {\bibinfo {author} {\bibfnamefont {S.}~\bibnamefont
  {Shi}}, \bibinfo {author} {\bibfnamefont {A.~L.}\ \bibnamefont {Wysocki}},\
  and\ \bibinfo {author} {\bibfnamefont {K.~D.}\ \bibnamefont {Belashchenko}},\
  }\bibfield  {title} {\bibinfo {title} {Magnetism of chromia from
  first-principles calculations},\ }\href
  {https://doi.org/10.1103/PhysRevB.79.104404} {\bibfield  {journal} {\bibinfo
  {journal} {Physical Review B—Condensed Matter and Materials Physics}\
  }\textbf {\bibinfo {volume} {79}},\ \bibinfo {pages} {104404} (\bibinfo
  {year} {2009})}\BibitemShut {NoStop}%
\bibitem [{\citenamefont {Astrov}(1961)}]{astrov1961magnetoelectric}%
  \BibitemOpen
  \bibfield  {author} {\bibinfo {author} {\bibfnamefont {D.}~\bibnamefont
  {Astrov}},\ }\bibfield  {title} {\bibinfo {title} {Magnetoelectric effect in
  chromium oxide},\ }\href {https://jetp.ras.ru/cgi-bin/dn/e_013_04_0729.pdf}
  {\bibfield  {journal} {\bibinfo  {journal} {Sov. Phys. JETP}\ }\textbf
  {\bibinfo {volume} {13}},\ \bibinfo {pages} {729} (\bibinfo {year}
  {1961})}\BibitemShut {NoStop}%
\bibitem [{\citenamefont {Botti}\ \emph {et~al.}(2005)\citenamefont {Botti},
  \citenamefont {Fourreau}, \citenamefont {Nguyen}, \citenamefont {Renault},
  \citenamefont {Sottile},\ and\ \citenamefont {Reining}}]{botti2005energy}%
  \BibitemOpen
  \bibfield  {author} {\bibinfo {author} {\bibfnamefont {S.}~\bibnamefont
  {Botti}}, \bibinfo {author} {\bibfnamefont {A.}~\bibnamefont {Fourreau}},
  \bibinfo {author} {\bibfnamefont {F.}~\bibnamefont {Nguyen}}, \bibinfo
  {author} {\bibfnamefont {Y.-O.}\ \bibnamefont {Renault}}, \bibinfo {author}
  {\bibfnamefont {F.}~\bibnamefont {Sottile}},\ and\ \bibinfo {author}
  {\bibfnamefont {L.}~\bibnamefont {Reining}},\ }\bibfield  {title} {\bibinfo
  {title} {Energy dependence of the exchange-correlation kernel of
  time-dependent density functional theory: A simple model for solids},\ }\href
  {https://doi.org/10.1103/PhysRevB.72.125203} {\bibfield  {journal} {\bibinfo
  {journal} {Physical Review B—Condensed Matter and Materials Physics}\
  }\textbf {\bibinfo {volume} {72}},\ \bibinfo {pages} {125203} (\bibinfo
  {year} {2005})}\BibitemShut {NoStop}%
\bibitem [{\citenamefont {Skovhus}\ and\ \citenamefont
  {Olsen}(2022)}]{skovhus2022magnons}%
  \BibitemOpen
  \bibfield  {author} {\bibinfo {author} {\bibfnamefont {T.}~\bibnamefont
  {Skovhus}}\ and\ \bibinfo {author} {\bibfnamefont {T.}~\bibnamefont
  {Olsen}},\ }\bibfield  {title} {\bibinfo {title} {Magnons in
  antiferromagnetic bcc {Cr} and {Cr$_2$O$_3$} from time-dependent density
  functional theory},\ }\href {https://doi.org/10.1103/PhysRevB.106.085131}
  {\bibfield  {journal} {\bibinfo  {journal} {Physical Review B}\ }\textbf
  {\bibinfo {volume} {106}},\ \bibinfo {pages} {085131} (\bibinfo {year}
  {2022})}\BibitemShut {NoStop}%
\bibitem [{\citenamefont {Szilva}\ \emph {et~al.}(2023)\citenamefont {Szilva},
  \citenamefont {Kvashnin}, \citenamefont {Stepanov}, \citenamefont
  {Nordstr{\"o}m}, \citenamefont {Eriksson}, \citenamefont {Lichtenstein},\
  and\ \citenamefont {Katsnelson}}]{szilva2023quantitative}%
  \BibitemOpen
  \bibfield  {author} {\bibinfo {author} {\bibfnamefont {A.}~\bibnamefont
  {Szilva}}, \bibinfo {author} {\bibfnamefont {Y.}~\bibnamefont {Kvashnin}},
  \bibinfo {author} {\bibfnamefont {E.~A.}\ \bibnamefont {Stepanov}}, \bibinfo
  {author} {\bibfnamefont {L.}~\bibnamefont {Nordstr{\"o}m}}, \bibinfo {author}
  {\bibfnamefont {O.}~\bibnamefont {Eriksson}}, \bibinfo {author}
  {\bibfnamefont {A.~I.}\ \bibnamefont {Lichtenstein}},\ and\ \bibinfo {author}
  {\bibfnamefont {M.~I.}\ \bibnamefont {Katsnelson}},\ }\bibfield  {title}
  {\bibinfo {title} {Quantitative theory of magnetic interactions in solids},\
  }\href {https://doi.org/10.1103/RevModPhys.95.035004} {\bibfield  {journal}
  {\bibinfo  {journal} {Reviews of Modern Physics}\ }\textbf {\bibinfo {volume}
  {95}},\ \bibinfo {pages} {035004} (\bibinfo {year} {2023})}\BibitemShut
  {NoStop}%
\bibitem [{\citenamefont {Esquembre-Ku\ifmmode \check{c}\else
  \v{c}\fi{}ukali\ifmmode~\acute{c}\else \'{c}\fi{}}\ \emph
  {et~al.}(2025)\citenamefont {Esquembre-Ku\ifmmode \check{c}\else
  \v{c}\fi{}ukali\ifmmode~\acute{c}\else \'{c}\fi{}}, \citenamefont {Le},
  \citenamefont {Garc\'{\i}a-Crist\'obal}, \citenamefont {Bernardi},
  \citenamefont {Sangalli},\ and\ \citenamefont
  {Molina-S\'anchez}}]{esquembre2025magnons}%
  \BibitemOpen
  \bibfield  {author} {\bibinfo {author} {\bibfnamefont {A.}~\bibnamefont
  {Esquembre-Ku\ifmmode \check{c}\else \v{c}\fi{}ukali\ifmmode~\acute{c}\else
  \'{c}\fi{}}}, \bibinfo {author} {\bibfnamefont {K.~B.}\ \bibnamefont {Le}},
  \bibinfo {author} {\bibfnamefont {A.}~\bibnamefont
  {Garc\'{\i}a-Crist\'obal}}, \bibinfo {author} {\bibfnamefont
  {M.}~\bibnamefont {Bernardi}}, \bibinfo {author} {\bibfnamefont
  {D.}~\bibnamefont {Sangalli}},\ and\ \bibinfo {author} {\bibfnamefont
  {A.}~\bibnamefont {Molina-S\'anchez}},\ }\bibfield  {title} {\bibinfo {title}
  {Magnons in chromium trihalides calculated with the ab initio
  {Bethe-Salpeter} equation},\ }\href {https://doi.org/10.1103/gbpw-zh1v}
  {\bibfield  {journal} {\bibinfo  {journal} {Phys. Rev. B}\ }\textbf {\bibinfo
  {volume} {112}},\ \bibinfo {pages} {184412} (\bibinfo {year}
  {2025})}\BibitemShut {NoStop}%
\bibitem [{\citenamefont {Olsen}(2021)}]{olsen2021unified}%
  \BibitemOpen
  \bibfield  {author} {\bibinfo {author} {\bibfnamefont {T.}~\bibnamefont
  {Olsen}},\ }\bibfield  {title} {\bibinfo {title} {Unified treatment of
  magnons and excitons in monolayer {CrI$_3$} from many-body perturbation
  theory},\ }\href {https://doi.org/10.1103/PhysRevLett.127.166402} {\bibfield
  {journal} {\bibinfo  {journal} {Physical Review Letters}\ }\textbf {\bibinfo
  {volume} {127}},\ \bibinfo {pages} {166402} (\bibinfo {year}
  {2021})}\BibitemShut {NoStop}%
\bibitem [{\citenamefont {M{\"u}ller}\ \emph {et~al.}(2016)\citenamefont
  {M{\"u}ller}, \citenamefont {Friedrich},\ and\ \citenamefont
  {Bl{\"u}gel}}]{muller2016acoustic}%
  \BibitemOpen
  \bibfield  {author} {\bibinfo {author} {\bibfnamefont {M.~C.}\ \bibnamefont
  {M{\"u}ller}}, \bibinfo {author} {\bibfnamefont {C.}~\bibnamefont
  {Friedrich}},\ and\ \bibinfo {author} {\bibfnamefont {S.}~\bibnamefont
  {Bl{\"u}gel}},\ }\bibfield  {title} {\bibinfo {title} {Acoustic magnons in
  the long-wavelength limit: investigating the {Goldstone} violation in
  many-body perturbation theory},\ }\href
  {https://doi.org/10.1103/PhysRevB.94.064433} {\bibfield  {journal} {\bibinfo
  {journal} {Physical Review B}\ }\textbf {\bibinfo {volume} {94}},\ \bibinfo
  {pages} {064433} (\bibinfo {year} {2016})}\BibitemShut {NoStop}%
\bibitem [{\citenamefont {Kshirsagar}\ and\ \citenamefont
  {Reichardt}(2025)}]{kshirsagar2025flipping}%
  \BibitemOpen
  \bibfield  {author} {\bibinfo {author} {\bibfnamefont {A.~R.}\ \bibnamefont
  {Kshirsagar}}\ and\ \bibinfo {author} {\bibfnamefont {S.}~\bibnamefont
  {Reichardt}},\ }\bibfield  {title} {\bibinfo {title} {Flipping of electronic
  spins in {BiFeO$_3$} via chiral d-d excitations},\ }\href
  {https://doi.org/10.1103/ht4g-7vb5} {\bibfield  {journal} {\bibinfo
  {journal} {Physical Review B}\ }\textbf {\bibinfo {volume} {112}},\ \bibinfo
  {pages} {L121111} (\bibinfo {year} {2025})}\BibitemShut {NoStop}%
\bibitem [{\citenamefont {Paleari}\ \emph {et~al.}(2025)\citenamefont
  {Paleari}, \citenamefont {Molina-Sánchez}, \citenamefont {Nalabothula},
  \citenamefont {Reho}, \citenamefont {Bonacci}, \citenamefont {Castelo},
  \citenamefont {Cervantes-Villanueva}, \citenamefont {Pionteck}, \citenamefont
  {Silvetti}, \citenamefont {Attaccalite},\ and\ \citenamefont {Pereira
  Coutada~Miranda}}]{paleari_2025_15012963}%
  \BibitemOpen
  \bibfield  {author} {\bibinfo {author} {\bibfnamefont {F.}~\bibnamefont
  {Paleari}}, \bibinfo {author} {\bibfnamefont {A.}~\bibnamefont
  {Molina-Sánchez}}, \bibinfo {author} {\bibfnamefont {M.}~\bibnamefont
  {Nalabothula}}, \bibinfo {author} {\bibfnamefont {R.}~\bibnamefont {Reho}},
  \bibinfo {author} {\bibfnamefont {M.}~\bibnamefont {Bonacci}}, \bibinfo
  {author} {\bibfnamefont {J.}~\bibnamefont {Castelo}}, \bibinfo {author}
  {\bibfnamefont {J.}~\bibnamefont {Cervantes-Villanueva}}, \bibinfo {author}
  {\bibfnamefont {M.}~\bibnamefont {Pionteck}}, \bibinfo {author}
  {\bibfnamefont {M.}~\bibnamefont {Silvetti}}, \bibinfo {author}
  {\bibfnamefont {C.}~\bibnamefont {Attaccalite}},\ and\ \bibinfo {author}
  {\bibfnamefont {H.}~\bibnamefont {Pereira Coutada~Miranda}},\ }\href
  {https://doi.org/10.5281/zenodo.15012963} {\bibinfo {title} {Yambopy}}
  (\bibinfo {year} {2025})\BibitemShut {NoStop}%
\bibitem [{\citenamefont {Hornreich}\ and\ \citenamefont
  {Shtrikman}(1968)}]{hornreich1968theory}%
  \BibitemOpen
  \bibfield  {author} {\bibinfo {author} {\bibfnamefont {R.}~\bibnamefont
  {Hornreich}}\ and\ \bibinfo {author} {\bibfnamefont {S.}~\bibnamefont
  {Shtrikman}},\ }\bibfield  {title} {\bibinfo {title} {Theory of gyrotropic
  birefringence},\ }\href {https://doi.org/10.1103/PhysRev.171.1065} {\bibfield
   {journal} {\bibinfo  {journal} {Physical Review}\ }\textbf {\bibinfo
  {volume} {171}},\ \bibinfo {pages} {1065} (\bibinfo {year}
  {1968})}\BibitemShut {NoStop}%
\bibitem [{\citenamefont {{\'I}niguez}(2008)}]{iniguez2008first}%
  \BibitemOpen
  \bibfield  {author} {\bibinfo {author} {\bibfnamefont {J.}~\bibnamefont
  {{\'I}niguez}},\ }\bibfield  {title} {\bibinfo {title} {First-principles
  approach to lattice-mediated magnetoelectric effects},\ }\href
  {https://doi.org/10.1103/PhysRevLett.101.117201} {\bibfield  {journal}
  {\bibinfo  {journal} {Physical Review Letters}\ }\textbf {\bibinfo {volume}
  {101}},\ \bibinfo {pages} {117201} (\bibinfo {year} {2008})}\BibitemShut
  {NoStop}%
\bibitem [{\citenamefont {Essenberger}\ \emph {et~al.}(2011)\citenamefont
  {Essenberger}, \citenamefont {Sharma}, \citenamefont {Dewhurst},
  \citenamefont {Bersier}, \citenamefont {Cricchio}, \citenamefont
  {Nordstr{\"o}m},\ and\ \citenamefont {Gross}}]{essenberger2011magnon}%
  \BibitemOpen
  \bibfield  {author} {\bibinfo {author} {\bibfnamefont {F.}~\bibnamefont
  {Essenberger}}, \bibinfo {author} {\bibfnamefont {S.}~\bibnamefont {Sharma}},
  \bibinfo {author} {\bibfnamefont {J.}~\bibnamefont {Dewhurst}}, \bibinfo
  {author} {\bibfnamefont {C.}~\bibnamefont {Bersier}}, \bibinfo {author}
  {\bibfnamefont {F.}~\bibnamefont {Cricchio}}, \bibinfo {author}
  {\bibfnamefont {L.}~\bibnamefont {Nordstr{\"o}m}},\ and\ \bibinfo {author}
  {\bibfnamefont {E.}~\bibnamefont {Gross}},\ }\bibfield  {title} {\bibinfo
  {title} {Magnon spectrum of transition-metal oxides: calculations including
  long-range magnetic interactions using the lsda+ u method},\ }\href
  {https://doi.org/10.1103/PhysRevB.84.174425} {\bibfield  {journal} {\bibinfo
  {journal} {Physical Review B—Condensed Matter and Materials Physics}\
  }\textbf {\bibinfo {volume} {84}},\ \bibinfo {pages} {174425} (\bibinfo
  {year} {2011})}\BibitemShut {NoStop}%
\bibitem [{\citenamefont {Timrov}\ \emph {et~al.}(2022)\citenamefont {Timrov},
  \citenamefont {Marzari},\ and\ \citenamefont {Cococcioni}}]{timrov2022hp}%
  \BibitemOpen
  \bibfield  {author} {\bibinfo {author} {\bibfnamefont {I.}~\bibnamefont
  {Timrov}}, \bibinfo {author} {\bibfnamefont {N.}~\bibnamefont {Marzari}},\
  and\ \bibinfo {author} {\bibfnamefont {M.}~\bibnamefont {Cococcioni}},\
  }\bibfield  {title} {\bibinfo {title} {Hp--a code for the calculation of
  hubbard parameters using density-functional perturbation theory},\ }\href
  {https://doi.org/10.1016/j.cpc.2022.108455} {\bibfield  {journal} {\bibinfo
  {journal} {Computer Physics Communications}\ }\textbf {\bibinfo {volume}
  {279}},\ \bibinfo {pages} {108455} (\bibinfo {year} {2022})}\BibitemShut
  {NoStop}%
\end{thebibliography}%

\clearpage
\onecolumngrid
\appendix
\renewcommand{\thesection}{S\arabic{section}}
\renewcommand{\theequation}{S\arabic{equation}}
\renewcommand{\thefigure}{S\arabic{figure}}
\renewcommand{\thetable}{S\arabic{table}}

\setcounter{section}{0}
\setcounter{equation}{0}
\setcounter{figure}{0}
\setcounter{table}{0}

\begin{center}
{\large\bfseries Supplemental Material for}\\[0.5em]
{\large\bfseries
``The Frequency-Dependent Spin Contribution to the
Magnetoelectric Tensor of Cr$_2$O$_3$:
A First-Principles Study''}
\end{center}

\vspace{1em}

\section{Relaxed geometry}
The final lattice parameters and atomic positions in crystal coordinates of Cr$_2$O$_3$ obtained with a variable-cell geometry optimization with GGA (PBE) are displayed in Table \ref{table:cr2o3structure}. The geometry relaxation was carried out using a collinear spin treatment and without spin-orbit coupling. The resulting unit cell magnetic moment, absolute magnetic moment, as well as the atomic magnetic moments and charges (obtained with a non-collinear GGA calculation with the relaxed geometry) are also reported in Table \ref{table:cr2o3structure}.
\begin{table}[h]
\centering
\caption{Relaxed geometry of Cr$_2$O$_3$ which was used in all calculations of the ME tensor in reduced units (ru) in terms of the lattice vectors. The total unit cell magnetic moment $\mathbf{M}$ and the absolute magnetic moment are shown along with the atomic magnetic moments and charges, obtained by integration over atomic spheres of radii $\SI{0.81}{\angstrom}$.}
\label{table:cr2o3structure}
\small

\[
a=b=c=\SI{5.420125}{\angstrom}, \qquad
\alpha=\beta=\gamma=\ang{54.516930}, \qquad
V=\SI{98.201644}{\angstrom^3}
\]
\[
M_x=M_y=M_z=\SI{0.00}{\mu_B}, \qquad
\text{Absolute magnetization}=\SI{11.80}{\mu_B}
\]

\begin{tabular}{
l
S[table-format=2.9]
S[table-format=1.9]
S[table-format=1.9]
S[table-format=1.9]
S[table-format=1.9]
S[table-format=2.6]
S[table-format=2.6]
}
\toprule
{Atom} &
{$x$, ru} &
{$y$, ru} &
{$z$, ru} &
{$m_x$ (\si{\mu_B})} &
{$m_y$ (\si{\mu_B})} &
{$m_z$ (\si{\mu_B})} &
{charge (\(-e\))} \\
\midrule
Cr1 & 0.002170 & 0.002170 & 0.002170 & 0.000000 & 0.000000 &  2.288774 & 10.859206 \\
Cr2 & 0.194173 & 0.194173 & 0.194173 & 0.000000 & 0.000000 & -2.290092 & 10.861849 \\
Cr3 & 0.502170 & 0.502170 & 0.502170 & 0.000000 & 0.000000 &  2.288774 & 10.859206 \\
Cr4 & 0.694173 & 0.694173 & 0.694173 & 0.000000 & 0.000000 & -2.290092 & 10.861849 \\
O1  & 0.788746 & 0.407352 & 0.098299 & 0.000000 & 0.000000 &  0.000205 &  5.483787 \\
O2  & 0.407352 & 0.098299 & 0.788746 & 0.000000 & 0.000000 &  0.000205 &  5.483787 \\
O3  & 0.098299 & 0.788746 & 0.407352 & 0.000000 & 0.000000 &  0.000205 &  5.483787 \\
O4  & 0.907352 & 0.288746 & 0.598299 & 0.000000 & 0.000000 &  0.000205 &  5.483787 \\
O5  & 0.288746 & 0.598299 & 0.907352 & 0.000000 & 0.000000 &  0.000205 &  5.483787 \\
O6  & 0.598299 & 0.907352 & 0.288746 & 0.000000 & 0.000000 &  0.000205 &  5.483787 \\
\bottomrule
\end{tabular}
\end{table}

\section{Dependence on the electronic-structure starting point}
\subsection{LDA and GGA comparison}
\begin{figure}[t!]
\begin{center}
\includegraphics[scale=1.0]{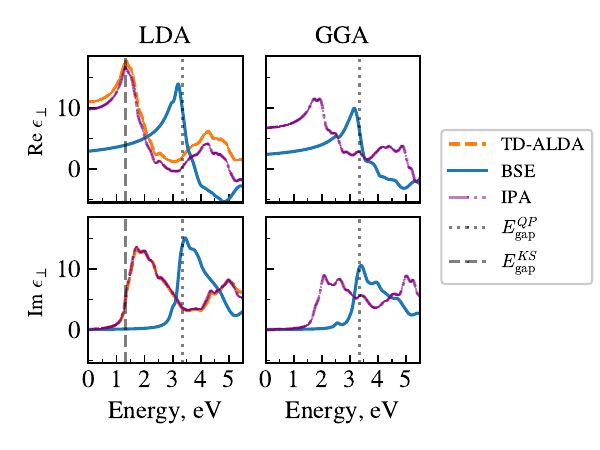}
    \caption{Comparisons between the transverse dielectric function obtained with the LDA (for TD-ALDA and BSE) and GGA (only for BSE) exchange correlation functionals. The independent-particle approximation (IPA) is shown for comparison.}
    \label{fig:epsilon_ALDA-BSE}
\end{center}
\end{figure}

\begin{figure}
\begin{center}
\begin{minipage}{0.49\linewidth}
\includegraphics[scale=1.0]{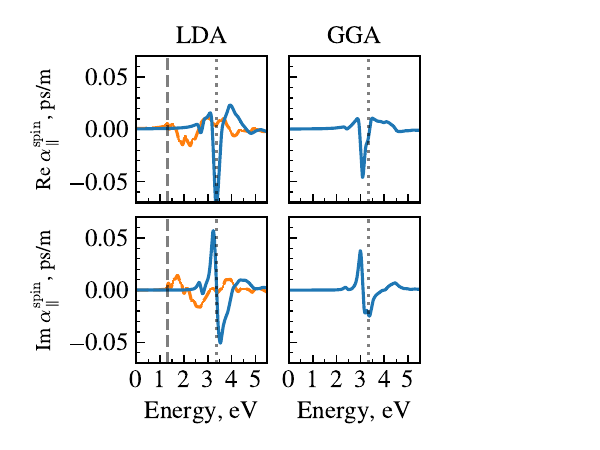}
\end{minipage}
\begin{minipage}{0.49\linewidth}
\hspace{-1.5cm}\includegraphics[scale=1.0]{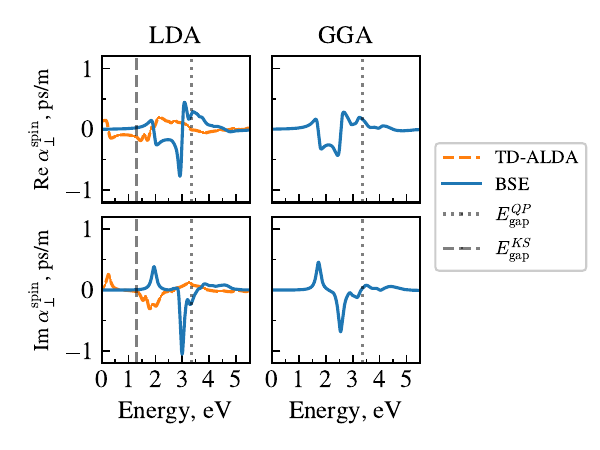}
\end{minipage}
    \caption{Comparisons between the longitudinal (left) spin ME response obtained with the LDA (for TD-ALDA and BSE) and GGA (only for BSE) exchange correlation functionals, and similarly for the transverse (right) spin ME response.}
    \label{fig:ME_ALDA-BSE}
\end{center}
\end{figure}

The LDA and GGA exchange--correlation functionals yielded similar but not identical BSE results. The major differences were the smaller optical oscillator strengths in GGA compared to LDA, as well as the relative position of the absorption onset, which occurs at lower energies in the case of GGA (Fig.~\ref{fig:epsilon_ALDA-BSE}). Due to the difference in exciton energies below the gap, the ME tensor is quantitatively different between LDA and GGA (Fig.~\ref{fig:ME_ALDA-BSE}), but similar in sign and shape, as is also seen in the comparison with the experimental rotation and ellipticity in the main text. Overall, while the choice of exchange–correlation functional leads to quantitative differences, the qualitative conclusions of this work remain unchanged.

\begin{figure*}[b]
\centering
\begin{minipage}[t]{0.49\linewidth}
\centering
\textbf{(a)}\\[0.5em]
\includegraphics[width=\linewidth]{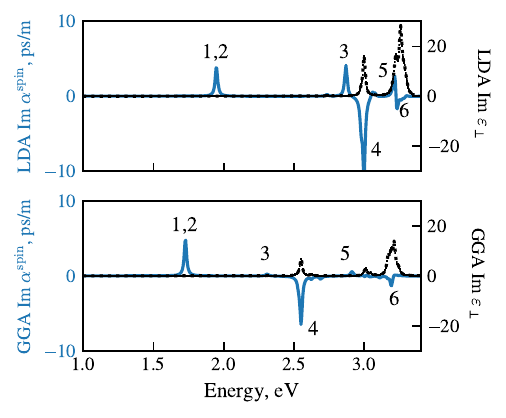}
\end{minipage}
\hfill
\begin{minipage}[t]{0.49\linewidth}
\centering
\textbf{(b)}\\[0.5em]
\footnotesize
\setlength{\tabcolsep}{5.0pt}
\begin{tabular}{c c c c c c}
Label & \(E\) (eV) & Deg. & \(\Delta S_z\) &
\(W_{d\to d}\) & \(\,R_\perp^\mathrm{spin}\) \\
\hline\hline
\multicolumn{6}{c}{LDA} \\
\hline
1 & 1.947 & 2 & \(-1\)    & 0.71 & \(-1.25\times10^{-2}\) \\
2 & 1.960 & 2 & \(+1\)    & 0.71 & \(-3.65\times10^{-4}\) \\
3 & 2.869 & 2 & \(\pm1\)  & 0.71 & \(-1.36\times10^{-2}\) \\
4 & 2.998 & 2 & \(0\)     & 0.69 & \(+3.29\times10^{-2}\) \\
5 & 3.219 & 1 & \(0\)     & 0.68 & \(-6.88\times10^{-3}\) \\
6 & 3.229 & 1 & \(0\)     & 0.68 & \(+6.47\times10^{-3}\) \\
\hline
\multicolumn{6}{c}{GGA} \\
\hline
1 & 1.728 & 2 & \(-1\)    & 0.69 & \(-1.57\times10^{-2}\) \\
2 & 1.739 & 2 & \(+1\)    & 0.69 & \(-3.30\times10^{-5}\) \\
3 & 2.309 & 4 & \(0\)     & 0.63 & \(-9.66\times10^{-4}\) \\
4 & 2.550 & 2 & \(\pm1\)  & 0.66 & \(+1.76\times10^{-2}\) \\
5 & 2.910 & 2 & \(\pm1\)  & 0.66 & \(+1.41\times10^{-3}\) \\
6 & 3.192 & 2 & \(-1\)    & 0.63 & \(+4.70\times10^{-3}\) \\
\end{tabular}
\end{minipage}

\caption{
Microscopic character of selected ME-active BSE poles obtained from LDA and GGA starting points. Panel (a) shows the transverse spin ME spectrum, together with the optical absorption, using a broadening $\eta=0.01$ eV. The numbered labels indicate the excitonic multiplets included in the analysis. Panel (b) lists the corresponding energy, degeneracy, spin character, Cr-$(d)\rightarrow$Cr-$(d)$ weight, and ME residuals $R_{\perp}^{\mathrm{spin}}$. The energies and $W_{d\rightarrow d}$ are averaged over each nearly degenerate multiplet, while the ME residual strengths are summed over the states in the multiplet.
}
\label{fig:si_exciton_character}
\end{figure*}

To characterize the microscopic nature of the selected ME-active BSE poles, we analyzed their spin and orbital composition. The excitonic spin character was obtained by projecting the spin operator onto the subspace spanned by each nearly degenerate excitonic multiplet. For a BSE exciton $|\lambda\rangle = \sum_{vc\mathbf{k}} A^\lambda_{vc\mathbf{k}}\, \hat a^\dagger_{c\mathbf{k}} \hat a_{v\mathbf{k}}|0\rangle$, the excitonic spin matrix was constructed as \cite{kshirsagar2025flipping}
\begin{equation}
\left(S_z^{\mathrm{exc}}\right)_{\lambda\lambda'} =
\sum_{\mathbf{k}vc}\left[
\sum_{c'}
\left(A^\lambda_{vc'\mathbf{k}}\right)^*
\langle c'\mathbf{k}|\hat S_z|c\mathbf{k}\rangle
A^{\lambda'}_{vc\mathbf{k}}
-
\sum_{v'}
\left(A^\lambda_{v'c\mathbf{k}}\right)^*
\langle v\mathbf{k}|\hat S_z|v'\mathbf{k}\rangle
A^{\lambda'}_{vc\mathbf{k}}
\right].
\label{eq:si_exciton_spin}
\end{equation}
The first term corresponds to the spin carried by the excited electron, while the second term accounts for the missing valence electron, i.e. the hole contribution. We then diagonalized \(S_z^{\mathrm{exc}}\) within each nearly degenerate multiplet, as implemented in Yambopy \cite{paleari_2025_15012963}. The resulting eigenvalues, reported as \(\Delta S_z\), therefore describe the dominant spin character of the excitonic multiplet. Values close to $\Delta S_z=\pm1$ indicate spin-flip excitations, while values close to $\Delta S_z=0$ indicate predominantly spin-conserving excitations. The orbital character is estimated by combining the BSE amplitudes with atomic projections of the underlying Kohn--Sham states,
\begin{equation}
W_{\alpha\to\beta} = \frac{\sum_{vc\mathbf{k}} |A^{\lambda}_{vc\mathbf{k}}|^2 P^{\alpha}_{v\mathbf{k}} P^{\beta}_{c\mathbf{k}}}{\sum_{vc\mathbf{k}} |A^{\lambda}_{vc\mathbf{k}}|^2},
\end{equation}
where $P^\alpha_{n\mathbf{k}}$ is the projected weight of the KS state $\ket{n\mathbf{k}}$ onto the subspace of atomic orbitals $\alpha$. The resulting character of the selected poles is summarized in Fig.~\ref{fig:si_exciton_character}. The numbered labels in panel (a) identify the excitonic multiplets listed in panel (b). Both LDA and GGA starting points yield predominantly Cr-$(d)\rightarrow$Cr-$(d)$ character for the selected ME-active poles, with $W_{d\rightarrow d}\simeq0.6-0.7$. However, the detailed spin character of individual poles is more sensitive to the starting point. In particular, the ordering and mixing of spin-flip and spin-conserving $d-d$-like excitations changes between LDA and GGA.


\FloatBarrier
\subsection{Effect of DFT$+U$ on BSE spectra}
The BSE calculations can be performed on top of a DFT ground-state where a Hubbard $U$ parameter has been included, as long as the Hubbard term in the Kohn-Sham Hamiltonian is accounted for in the calculation of the optical dipole elements. This can be done by utilizing a covariant derivative approach as described in Ref.~\cite{sangalli2019many}. The Hubbard correction can be applied in order to provide a better starting point with a more realistic description of the localization of $d$-electron states. The inclusion of $U$ affects both occupied as well as unoccupied Kohn-Sham orbitals, with the general trend of shifting the bands of $d$-orbital character away from the Fermi level \cite{essenberger2011magnon}. The determination of the value of $U$ is often not straightforward, and it is sometimes treated as an empirical fitting parameter. There have been developments in determining it from first-principles methods, however \cite{timrov2022hp}.

In order to study the effect of a Hubbard $U$ correction, we re-calculated the BSE absorption and ME spectra with GGA$+U$. We used the same set of computational parameters as in the main text apart from the k-mesh, for which we used a finer $8\times 8\times 8$ unshifted grid as the coarse grid in order to guarantee an accurate numerical covariant derivative. We considered $U=5$ eV, which has been previously used in other computational studies \cite{lebreau2014structural}. The Hubbard correction is, however, dependent on the atomic projections and pseudopotential and the determination of a more appropriate value of $U$ would require a more careful study. Here we only consider the general effect and the robustness of the results against its inclusion. 

\begin{figure}[h!]
\begin{center}
    \includegraphics[scale=0.9]{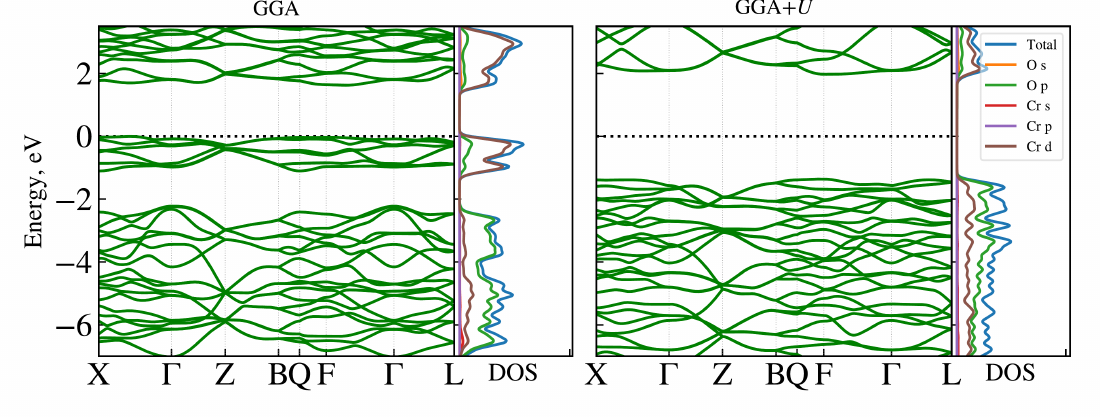}
    \caption{The GGA bands (left) and GGA$+U$ bands (right) for Cr$_2$O$_3$, using a value of $U=5$ eV. As expected, the inclusion of $U$ moves the bands with a high Cr-$d$ character away from the Fermi level, leading to a mixing of the Cr-$d$ and O-$p$ valence states.}
    \label{fig:dft+u_bands}
\end{center}
\end{figure}

\begin{figure}[t!]
\begin{center}
    \includegraphics[scale=1.0]{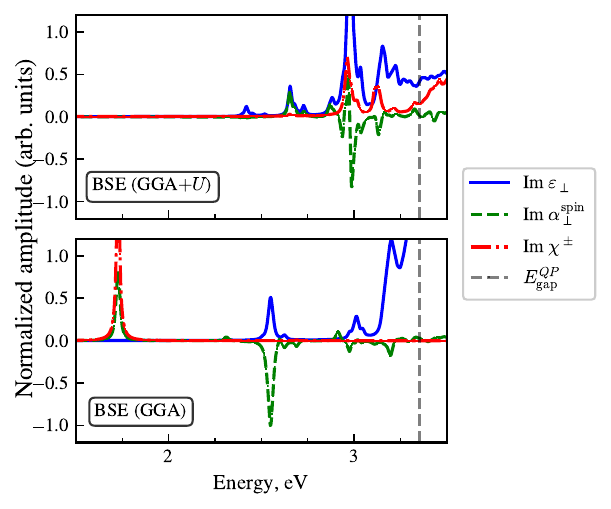}
    \caption{The imaginary parts of the transverse dielectric tensor component ($\operatorname{Im}\varepsilon_\perp$), spin susceptibility ($\operatorname{Im}\chi^\pm$), and transverse spin ME tensor element ($\operatorname{Im}\alpha_\perp^{\mathrm{spin}}$), obtained with BSE on top of GGA+$U$ (top, $U=5$ eV) and GGA (bottom) ground states. The quasiparticle gap, fixed at 3.4 eV in both calculations, is indicated by the vertical dashed line. The broadening is $\eta=0.01$ eV.}
    \label{fig:dft+u_bse}
\end{center}
\end{figure}

As can be seen in Fig.~\ref{fig:dft+u_bands}, the inclusion of $U$ has the effect of opening up the band gap by shifting the bands with high Cr-$d$ orbital character away from the Fermi level. This leads to the valence band edge being composed of a mixture of Cr-$d$ and O-$p$ bands, instead of the more isolated Cr-$d$ valence band edge in the Hubbard-free case. This affects both the optical and spin-dependent matrix elements entering the BSE calculation, leading to significant changes in the BSE spectra as shown in Fig.~\ref{fig:dft+u_bse}. Although the quasiparticle gap is kept fixed at the same value (3.4 eV) in both calculations via a scissor correction, we still observe a shift towards higher energies and a redistribution of spectral weight in the BSE spectra when including the Hubbard $U$. This indicates that the changes arise from modifications of the underlying wavefunctions and screening entering the BSE calculation. The transverse ME tensor element continues to exhibit a clear excitonic resonance structure upon the inclusion of the Hubbard $U$. In particular, the sign of the ME response at the dominant low-energy feature, associated with the peak of the spin susceptibility, is preserved between GGA and GGA$+U$ starting points. Similarly, the higher energy excitonic peak exhibits the same sign in both cases. However, the main features of the spin susceptibility are shifted towards higher energies, approaching the energy range of the optical excitonic resonances, whereas they were well separated from the optical resonances in the Hubbard-free case. This highlights the sensitivity of the spin-dependent response to the degree of localization of the Cr-$d$ states.

\FloatBarrier
\section{Convergence tests}
\begin{figure}[h!]
\begin{center}
\begin{minipage}{0.49\linewidth}
    \includegraphics[scale=1.2]{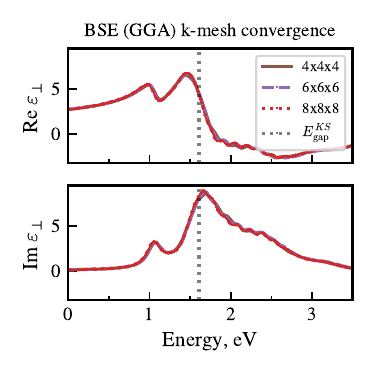}
\end{minipage}
\begin{minipage}{0.49\linewidth}
    \includegraphics[scale=1.2]{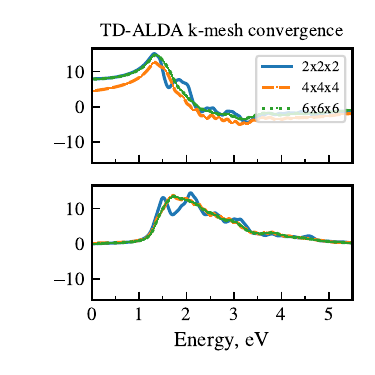}
\end{minipage}
\end{center}
\caption{Convergence of the k-mesh for BSE (GGA), left, and TD-ALDA, right, with respect to the dielectric function. The k-mesh was always shifted by half a grid step in each direction. The BSE calculation made use of a double-grid method with a fine mesh of $16\times 16\times 16$, and a smaller screening cutoff of 2 Ry than the production runs. For the k-mesh convergence tests no rigid scissor correction was applied, since the scissor primarily produces a rigid shift of the excitation energies and does not affect the convergence behavior of the spectral shape or relative peak distribution. The scissor correction is applied in the band-convergence tests shown below and in the production calculations discussed in the main text.}
\label{fig:kmesh_convergence}
\begin{center}
\begin{minipage}{0.49\linewidth}
    \includegraphics[scale=1.2]{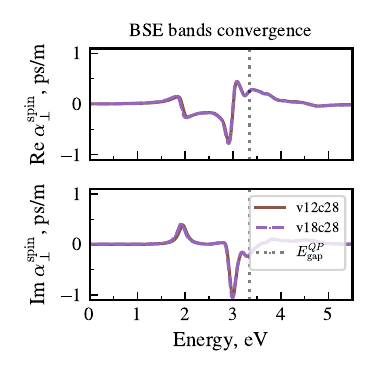}
\end{minipage}
\begin{minipage}{0.49\linewidth}
    \includegraphics[scale=1.2]{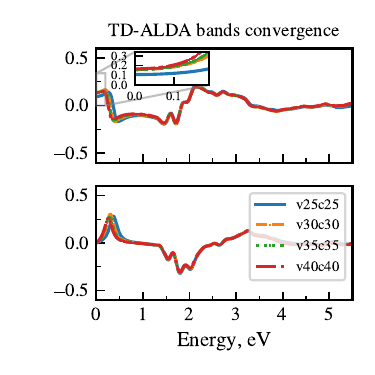}
\end{minipage}
\end{center}
\caption{Convergence of the number of bands included in the kernel for BSE (LDA), left, and TD-ALDA, right, with respect to the ME tensor. The legend indicates the number of valence bands (v) and conduction bands (c). The broadening used was $\eta=0.1$ eV, except in the inset where a broadening of $\eta=0.01$ eV was used.}
\label{fig:bands_convergence}
\end{figure}

The convergence of the TD-ALDA and BSE spectra with respect to k-points and bands included in the kernel is shown in Fig.~\ref{fig:kmesh_convergence} and \ref{fig:bands_convergence}. During the convergence tests, we observed that the ME tensor converged faster with respect to the k-point density than the dielectric tensor, whereas convergence with respect to the number of bands was slower. The convergence of the lowest energy peak with respect to the bands proved to be the main challenge for both BSE and TD-ALDA calculations. The necessity of including many bands in the kernel for the convergence of low-energy states has been mentioned in other $GW$--BSE approaches to magnon calculations \cite{esquembre2025magnons}.

\FloatBarrier

\end{document}